\documentclass[aps,epsfig]{revtex4}

\usepackage{epsfig}

\begin{document}

\newcommand{\be}{\begin{equation}}
\newcommand{\ee}{\end{equation}}
\newcommand{\bea}{\begin{eqnarray}}
\newcommand{\eea}{\end{eqnarray}}

%\draft
%\twocolumn[\hsize\textwidth\columnwidth\hsize\csname@twocolumnfalse\endcsname

\title{\bf Viscosity in molecular dynamics with periodic boundary conditions}

\author{S. Viscardy and P. Gaspard\\
{\em Center for Nonlinear Phenomena and Complex Systems,}\\
{\em Universit\'e Libre de Bruxelles,}
{\em Campus Plaine, Code Postal 231, B-1050 Brussels, Belgium}\\}
%\address{}

\date{\today}

\begin{abstract}
We report a study of viscosity by the method of Helfand moment in systems
with periodic boundary conditions.  We propose a new definition of 
Helfand moment which takes into account the minimum image convention 
used in molecular dynamics with periodic boundary conditions.  
Our Helfand-moment method is equivalent to the method based on the 
Green-Kubo formula and is not affected by ambiguities due to the 
periodic boundary conditions.  Moreover, in hard-ball systems, 
our method is equivalent to the one developed by 
B. J. Alder, D. M. Gass, and T. E. Wainwright 
[{\it J. Chem. Phys.} {\bf 53}, 3813 (1970)]. 
We apply and verify our method in a fluid composed 
of $N\geq 2$ hard disks in elastic collisions.   
We show that the viscosity coefficients already take values
in good agreement with Enskog's theory for $N=2$ hard disks
in a hexagonal geometry.
\end{abstract}

\maketitle

\vskip 0.3 cm

\hspace*{16mm}{PACS numbers: 02.70.Ns; 05.60.-k; 05.20.Dd}

\section{Introduction}

Viscosity is the fundamental mechanism of dissipation of momentum in a fluid.
Viscosity is described at the macroscopic level by the Navier-Stokes equations
which are the equations of balance of momentum in a fluid.  At the
microscopic level, viscosity arises because of a transfer of momentum 
between fluid layers moving at different
velocities as already explained by Maxwell thanks to kinetic theory.

In the fifties, Green, Kubo, Mori, and others provided an explanation
of all the transport properties in terms of time-dependent statistical 
correlations of microscopic currents
associated with each transport property \cite{green51,green60,kubo57,mori58}.
They showed that the transport coefficients are given 
as the time integrals of the time autocorrelation functions of
the microscopic currents, yielding the famous Green-Kubo formulas.  
Thereafter, Helfand showed in the early sixties that 
the transport coefficients can be expressed by Einstein-like
formulas in terms of moments -- the so-called Helfand moments -- 
which are the time integrals of the microscopic currents
\cite{helf}.

These new methods by Green, Kubo, Mori, Helfand, and others have been applied
to the computation of transport properties by molecular-dynamics simulations, 
in particular, by Alder {\it et al.} \cite{alder}.  
In molecular-dynamics simulations, the system is necessarily composed 
of a finite number of particles which are usually moving in a domain 
defined with periodic boundary conditions in order to simulate 
the bulk properties.  The periodic boundary conditions (p.b.c.)
usually considered in molecular dynamics are based on the so-called {\it
minimum image convention} according to which interaction 
should occur between pairs of particles separated by the minimum
distance among the infinitely many images of the particles allowed by the
p.b.c..  In molecular-dynamics simulations, the minimum image convention 
plays a fundamental role to define
the microscopic current entering the Green-Kubo formula.

We may wonder if the Helfand-moment method could be applied
to molecular dynamics simulations with p.b.c..  The advantage of the
Helfand-moment method is that it expresses the transport
coefficients by Einstein-like formulas, directly showing their positivity.
Moreover, this method is very efficient because it is
based on a straightforward accumulation which is numerically robust.
Actually, it is a Helfand-moment method which has been 
numerically implemented by Alder {\it et al.} for viscosity in
hard-ball fluids \cite{alder}.  Several other implementations of the
Helfand-moment method have been considered and discussed in the literature
\cite{CD,A,erpenbeck}. However, the implementation of this method
for systems subject to p.b.c. other than hard-ball fluids seems to remain 
ambiguous as reported by Erpenbeck in Ref. \cite{erpenbeck}.

The purpose of the present paper is to propose a Helfand-moment method
which is appropriate for molecular dynamics simulations with p.b.c. and which
is strictly equivalent to calculations with the Green-Kubo formula.
For this purpose, we show the need to take into account the minimum image
convention.  In this way, we are able to obtain a Helfand moment giving
viscosity thanks to an Einstein-like formula in molecular dynamics
with p.b.c..  The so-obtained value of viscosity is in full agreement with the value
of the Green-Kubo formula and also with the value obtained by Alder {\it et al.}
\cite{alder}.

Our method is applied to the hard-disk fluid.  We study in detail the simple model
composed of two hard disks in elastic collisions in a domain defined by p.b.c.. 
Due to the defocusing character of the disks,
this model is chaotic. Bunimovich and Spohn have demonstrated that the
viscosity already exists in this system with 
only two particles \cite{buni-spohn}. The model they studied
is defined with p.b.c. in a square geometry. It presents a fluid and
a solid phases which are separated by a phase transition.
The problems presented by the model in a square geometry are that:
$(i)$ the viscosity exists only in the solid phase; \break
$(ii)$ the viscosity tensor which is of fourth order is
\textit{anisotropic} on a square lattice. In the present work, 
we solve these problems by considering a hexagonal geometry.
Indeed, in the hexagonal geometry, the fourth-order viscosity tensor is
isotropic and we can proof the existence of viscosity 
already in the fluid phase.

Furthermore, we apply our method to systems containing more and more hard disks.
We show that the values of the shear viscosity obtained by our
Helfand-moment method are in good agreement with Enskog's theory, already
for the fluid of two-hard disks.

The paper is organized as follows. In Sec. \ref{sec.helfand}, we derive our
expression of the Helfand moment for the viscosity tensor to be 
of application in molecular-dynamics simulations
with p.b.c.. In Sec. \ref{sec.2disks}, we describe the model of 
two hard disks in the hexagonal and square geometries. In Sec.
\ref{sec.2disks.prop}, we study different properties of the 
model like the mean free path and the hydrostatic pressure, in
particular, across the fluid-solid phase transition.  In Sec.
\ref{sec.2disks.visc}, the Helfand-moment method is applied to 
the two-disk model to calculate the shear and bulk
viscosities.  We show how the fluid-solid phase transition affects the
viscosities in this model.  In Sec. \ref{sec.Ndisks.visc}, we extend
our results to systems with $N=4,8,12,...,40$ hard disks.
We show that the shear viscosity already takes a value in good
agreement with Enskog's theory in the two-hard-disk system.
Our results are discussed and 
conclusions are drawn in Sec. \ref{sec.conclusions}.

%%%%%%%%%%%%%%%%%%%%%%%%%%%%%%%%%%%%%%%%%%%%%%%%%%%%%%%%%%%%%%%%%%%%%%%%%%%%%%%%
%%%%%%%%%%%%%%%%%%%%%%%%%%%%%%%%%%%%%%%%%%%%%%%%%%%%%%%%%%%%%%%%%%%%%%%%%%%%%%%%

\section{Hydrodynamics, Helfand moment, and viscosity}
\label{sec.helfand}

\subsection{Viscosity and hydrodynamics}

The hydrodynamic theory provides us with the equations
of motion for the conserved quantities in a fluid. In
particular, the local conservation of momentum is expressed by
the well-known \textit{Navier-Stokes equations}
\cite{landau-fluid}:

\begin{equation}
\frac{\partial \rho v_{i}}{\partial t}
=-\frac{\partial \Pi_{ij}}{\partial r_{j}} \; ,
\end{equation}
where 

\begin{equation}
\Pi_{ij}= \rho \, v_{i}v_{j} + P \; \delta_{ij} - \sigma'_{ij}\; ,
\end{equation}
is the momentum flux density tensor, $\rho v_i$ is the momentum density, 
$P$ the hydrostatic pressure, and $\sigma'_{ij}$ the viscosity stress tensor.  
This last tensor takes into account the internal friction occurring in a fluid
when different parts of the fluid move with different velocities.
Therefore, $\sigma'_{ij}$ has to be proportional to the space
derivatives of the velocities:

\begin{equation}
\sigma'_{ij}=\eta_{ij,kl} \; \frac{\partial v_{k}}{\partial r_{l}} \; ,
\end{equation}
where $\eta_{ij,kl}$ is the \textit{viscosity tensor}.

For isotropic systems, the theory of Cartesian tensors
shows that the basic isotropic tensor is the Kronecker tensor $\delta_{ij}$
and that all the isotropic tensors of even order
can be written like a sum of products  of tensors $\delta_{ij}$ \cite{aris}:

\begin{equation}
\eta_{ij,kl}=a \; \delta_{ij} \;\delta_{kl}
+ b \; \delta_{ik} \;\delta_{jl} + c \; \delta_{jk} \;\delta_{il}\; ,
\end{equation}
where $a,b$ and $c$ are scalars. Since the viscosity stress tensor is symmetric
$\sigma'_{ij}=\sigma'_{ji}$, only two of these coefficients are independent
because $b=c$.
After a rearrangement, we have:

\begin{equation}
\sigma'_{ij}=\eta \left ( \frac{\partial v_{i}}{\partial r_{j}}
+ \frac{\partial v_{j}}{\partial r_{i}} - \frac{2}{d}\;
\delta_{ij}\, \frac{\partial v_{l}}{\partial r_{l}} \right )
+ \zeta \; \delta_{ij} \,  \frac{\partial v_{l}}{\partial r_{l}}\; ,
\end{equation}
for a $d$-dimensional system.  The coefficients $\eta=b$ and
$\zeta=a+(2/d)b$ are respectively the
\textit{shear and bulk viscosities} and they can be expressed in terms of the
elements of the
fourth-order viscosity tensor as:
\be
\cases{\displaystyle \eta=\eta_{xy,xy}\; , \cr
\displaystyle \zeta = \frac{1}{d}\; \eta_{xx,xx} + \frac{d-1}{d}\;
\eta_{xx,yy}\; . \cr}
\ee

%%%%%%%%%%%%%%%%%%%%%%%%%%%%%%%%%%%%%%%%%%%%%%%%%%%%%%%%%%%%%%%%%%%%%%%%%%%%%%%%

\subsection{The Green-Kubo formula in molecular dynamics with p.b.c.}

Several techniques have been developed during
the last century to evaluate the transport coefficients.
One of the most important methods was established by
Green, Kubo, and Mori \cite{green51,green60,kubo57,mori58}.
It consists in having a relation between each
transport coefficient and the autocorrelation
function of the associated flux or microscopic current. In our case (see
Appendix \ref{appendixA}), we have:

\begin{equation}
\eta_{ij,kl}=\frac{\beta}{V}\int_{0}^{\infty} \left[ \langle
J_{ij}(0)\;J_{kl}(t)
\rangle - \langle J_{ij}\rangle \langle J_{kl}\rangle \right] \; dt \; ,
\label{autocorrel}
\end{equation}
with the microscopic current:

\begin{eqnarray}
J_{ij}&=& \sum_{a=1}^{N}\frac{p_{ai} \;
p_{aj}}{m} + \frac{1}{2} \sum_{a\neq b=1}^{N}
F_{i}({\mathbf r}_a - {\mathbf r}_b) \; \left (r_{aj} -  r_{bj} \right ) \; ,
\label{flux}
\end{eqnarray}
${\mathbf p}_a$ and ${\mathbf r}_a$ being the momentum and the
position of the $a^{th}$ particle, while ${\mathbf F}({\mathbf r}_a -
{\mathbf r}_b)$
is the force between particles $a$ and $b$.
In Eq. (\ref{autocorrel}) the average $\langle\cdot\rangle$ is performed
with respect to the equilibrium state.  We notice that, for the
microcanonical state,
\be
\beta=\frac{1}{k_{\rm B}T}\; \frac{N}{N-1} \; ,
\ee
(see Appendix \ref{appendixA}).

A very important point is that, in a system with p.b.c. as considered in
molecular-dynamics simulations, the difference of positions
${\bf r}_a-{\bf r}_b$ must satisfy the {\it minimum image convention} that
\be
\vert r_{aj} -r_{bj}\vert \leq \frac{L}{2} \; , \qquad {\rm for} \ \
j=1,...,d\; ,
\ee
for a cubic geometry.  More generally, the difference of positions must remain
within a unit cell of the Bravais lattice used to define the p.b.c..  With
p.b.c., there is indeed an infinite lattice of images of each particle.  All
these images move in parallel. If the force has a finite range the particle
$a$ interacts only
with the particles $b$ within its interaction range.  The force field ${\bf
F}({\bf r})$ has a
finite range of interaction beyond which it vanishes.  The interaction
range is supposed to be
smaller than the size $L$ of the box containing all the particles.  It is
important to notice
that we do not suppose here that the force field is periodic.  In order to
define a dynamics which
is periodic in the box of size $L$ the positions should jump in order to
satisfy the minimum image
convention.  As a consequence of this assumption, the positions and momenta
used to calculate the
viscosity by the Green-Kubo method actually obey modified Newton's equations
\be
\cases{\displaystyle
\frac{d{\bf r}_a}{dt} = \frac{{\bf p}_a}{m} + \sum_s \Delta{\bf
r}_{a}^{(s)} \; \delta(t-t_s)\; ,\cr
\displaystyle\frac{d{\bf p}_a}{dt} = \sum_{b (\neq a)} {\bf F}({\bf
r}_a-{\bf r}_b)  \; ,}
\label{Newton}
\ee
where $\Delta{\bf r}_{a}^{(s)}$ is the jump of the particle $a$ at time
$t_s$ in order to satisfy the minimum image convention.  Moreover, we
impose that the particle No. 1
does not jump. To satisfy these conditions, the jumps at the time $t_s$
when $\vert
r_{aj}(t_s)-r_{bj}(t_s)\vert = L/2$ can be given by
\be
{\rm for}\ \ a<b: \quad \cases{ \Delta r_{aj}^{(s)} = 0 \; , \cr
\Delta r_{bj}^{(s)} =  \varepsilon L  \; ,\cr
\Delta r_{cj}^{(s)} =0 \; , \qquad {\rm for}\ \ c\neq a,b \; ,\cr
\Delta r_{dk}^{(s)} = 0\; , \qquad {\rm for}\ \ k\neq j \quad{\rm and}\ \
\forall d \; ,\cr}
\ee
with $\varepsilon ={\rm sgn}[p_{aj}(t_s)-p_{bj}(t_s)]$.  The
modified Newton equations (\ref{Newton}) define a dynamics which is
periodic on the torus of the
relative coordinates ${\bf r}_a-{\bf r}_1$ because the jumps of the
relative coordinates are
vectors of the Bravais cubic lattice: $\Delta r_{aj}^{(s)}-\Delta
r_{1j}^{(s)} =0,\pm L$, while
the momenta ${\bf p}_a$ remain functions of the time without singularities
worst than
discontinuities.  We notice that modified Newton's equations (\ref{Newton})
conserve energy, total momentum and preserve phase-space volumes
(Liouville's theorem).

%%%%%%%%%%%%%%%%%%%%%%%%%%%%%%%%%%%%%%%%%%%%%%%%%%%%%%%%%%%%%%%%%%%%%%%%%%%%%%%%

\subsection{Helfand moment for molecular dynamics with p.b.c.}

In the sixties, Helfand has derived quantities associated with the
different transport processes,
in particular for the viscosities \cite{helf}. These new quantities
$G_{ij}(t)$ are
such that we can obtain an Einstein-like relation for each transport
coefficient. For the shear
viscosity coefficient, we have:

\begin{equation}
\eta=\lim_{t \to \infty} \frac{\beta}{2t V}\; \left \langle \left[G_{xy}
(t)-G_{xy} (0)
\right]^{2}\right \rangle \; .
\label{Einstein.shear.viscosity}
\end{equation}
More generally, we can define such a relation for each element of the
viscosity tensor:
\begin{equation}
\eta_{ij,kl}=\lim_{t \to \infty} \frac{\beta}{2t V}\; \left[ \langle
G_{ij}(t)G_{kl}(t)
\rangle - \langle G_{ij}(t) \rangle \langle G_{kl}(t) \rangle \right]
\label{Einstein}
\end{equation}
if we take $G_{ij}(0)=0$.  The \textit{Helfand moment} $G_{ij}(t)$ is
defined as the integral of
the microscopic current appearing in the Green-Kubo relation:
\be
G_{ij}(t) = G_{ij}(0) + \int_0^t J_{ij}(\tau) \; d\tau \; .
\label{helfand-integral}
\ee
As a consequence of the definition (\ref{helfand-integral}), the
Einstein-Helfand formula (\ref{Einstein}) is equivalent to the Green-Kubo
formula (\ref{autocorrel}), as proved in Appendix \ref{appendixB}.  In a
system of $N$ particles on a torus and satisfying the minimum image
convention, we can integrate the current (\ref{flux}) with modified Newton's
equations (\ref{Newton}) to get:
\be
G_{ij}(t) = \sum_{a=1}^N p_{ai}(t) \; r_{aj}(t) - \sum_{a=1}^N \sum_s
p_{ai}^{(s)}
\; \Delta r_{aj}^{(s)} \; \theta(t-t_s) \; ,
\label{helfand-torus}
\ee
where $G_{ij}(0)=0$, $p_{ai}^{(s)}=p_{ai}(t_s)$ and $\theta(t-t_s)$ is the
{\it Heaviside step
function} at the time $t_s$ of the jump $s$:
\be
\theta(t-t_s) = \cases{ 1 \; , \quad {\rm for} \ \ t>t_s \; ,\cr
0 \; , \quad {\rm for} \ \ t<t_s \; .\cr}
\ee
The expression (\ref{helfand-torus}) which we propose in the present paper
can be used to obtain the viscosity coefficients thanks to the
Einstein-like formulas (\ref{Einstein}) in a molecular dynamics defined on the
torus.  We emphasize that the expression (\ref{helfand-torus})
may apply to systems of particles interacting with a smooth potential under the
condition that the range is finite, or to systems of hard balls in elastic
collisions.  We show in Appendix \ref{appendixC} that the hydrostatic pressure
can also be written in terms of the Helfand moment (\ref{helfand-torus}).

Our Helfand-moment method has several theoretical and numerical advantages:
(i) It is strictly equivalent to the Green-Kubo method.  (ii) The
Einstein-like formula (\ref{Einstein.shear.viscosity}) or (\ref{Einstein})
directly show the positivity of the viscosity coefficient or viscosity tensor
because $t$, $\beta$, and $V$ are positive.  Moreover, the Helfand moments
directly obey central limit theorems, expressing the Gaussian character
of the dynamical fluctuations in systems with finite viscosity.
(iii) Thanks to our expression (\ref{helfand-torus}) of the Helfand moment,
the viscosity coefficients are given by a straightforward accumulation
over the successive jumps $s$.  For a given system with $N$ particles,
numerical convergence can be reached in the limit of an arbitrarily large
number of jumps $s$, under conditions of existence of the viscosity coefficients.

By defining the Helfand moment as the integral (\ref{helfand-integral}) of the
microscopic current for a system with minimum image convention, we obtain
the expression (\ref{helfand-torus}) which can be used to directly calculate
$\Delta G_{ij}(t)=G_{ij}(t)-G_{ij}(0)$ for the Einstein-Helfand relation,
remaining consistent with the requirements imposed by the periodic boundary
conditions and with the Green-Kubo formula for a system satisfying the minimum image
convention. 

%%%%%%%%%%%%%%%%%%%%%%%%%%%%%%%%%%%%%%%%%%%%%%%%%%%%%%%%%%%%%%%%%%%%%%%%%%%%%%%%

\subsection{Comparison with other methods}
\label{subsec.Helfand-hard-ball}

In the seventies, Alder et al. \cite{alder} calculated the viscosity
coefficients of hard-ball
systems with Einstein-like formulas based on expressions for Helfand
moments which are specific to
hard-ball systems. Instead of adding a new quantity to the Helfand
moment at each passage through the boundaries of the minimum image
convention as in Eq.
(\ref{helfand-torus}), their expression takes into account only the elastic
collisions between the
hard balls.  The Helfand moment can be obtained by direct integration of
the microscopic current
according to Eq. (\ref{helfand-integral}) with $G_{ij}(0)=0$:
\begin{eqnarray}
G_{ij}(t) & = &  \int_{0}^{t} d\tau \; J_{ij}(\tau)\\
& = & \int_{0}^{t} d\tau \left [ \sum_{a=1}^{N} \frac{p_{ai}p_{aj}}{m}
+\frac{1}{2} \sum_{a\neq b}^{}
F_{i}({\bf r}_{a} - {\bf r}_{b}) \; (r_{aj} - r_{bj})\right] \, .
\end{eqnarray}
Between the collisions the trajectory is a straight line and the
particle velocities change only at each collision. Therefore, the first
term in the integral, the
kinetic term, is constant during two successive collisions and changes only
at the collisions.
The second term, the potential term, vanishes between two successive
collisions and contributes
only at collisions.  Indeed, for a hard-ball potential, the forces between
the particles $a$ and
$b$ colliding at the time $t_c$ of the collision $c$ can be written in
terms of the
change $\Delta{\bf p}_{a}^{(c)}={\bf p}_{a}(t_c+\epsilon)-{\bf
p}_{a}(t_c-\epsilon)$ of momentum
of the particle $a$ at the collision $c$ as
\be
\cases{
{\bf F}\left({\bf r}_a- {\bf r}_b \right) = +\Delta{\bf p}_{a}^{(c)} \;
\delta(t-t_{c}) \; ,\cr
{\bf F}\left({\bf r}_b- {\bf r}_a \right) = -\Delta{\bf p}_{a}^{(c)} \;
\delta(t-t_{c}) \; ,\cr}
\label{Dp}
\ee
for $t_c-\epsilon < t < t_c+\epsilon$, because $\Delta{\bf
p}_{b}^{(c)}=-\Delta {\bf
p}_{a}^{(c)}$.  The forces with the other particles which are not engaged
in the collision vanish.
Therefore, we obtain:
\begin{equation}
G_{ij}(t)= \sum_{(c-1,c)} \left( \sum_{a=1}^{N}\frac{p_{ai}
p_{aj}}{m}\right)_{(c-1,c)} \Delta
t_{c-1,c} + \sum_{c} \Delta p_{ai}^{(c)} \; r_{abj}^{(c)} \; \theta(t-t_c) \; ,
\label{alder-expression}
\end{equation}
where, in the first term, $\Delta t_{c-1,c}$ is the time of flight between the
collisions $c-1$ and $c$ during which the momenta remain constant and, in
the second term, $a$ and $b$
denote the particles interacting at the collision $c$ and
$r_{abj}^{(c)}=r_{aj}(t_c) - r_{bj}(t_c)$.
The first sum runs over the intercollisional free flights $(c-1,c)$ between
the initial time $t=0$ and
the current time $t$, while the second sum runs over the collisions
occurring between the
time $t=0$ and $t$.  If $C$ denotes the last collision before the current
time $t$, we notice that the
last term of the first sum is $\Delta t_{C,C+1}=t-t_C$.  Hence, if we
differentiate Eq.
(\ref{alder-expression}) with respect to time and use (\ref{Dp}) we recover
the microscopic current
(\ref{flux}).  Therefore, the expression (\ref{alder-expression}) is
equivalent to our expression
(\ref{helfand-torus}) in the case of hard-ball systems.  However, our
expression (\ref{helfand-torus})
extends to systems with a smooth interaction potential.

A comment is here in order about another method 
which has been considered and discussed  
in the literature \cite{CD,A,erpenbeck}.
This other method implements an expression 
printed in the middle of a presentation
given in Ref. \cite{McQuarrie} for the calculation
of shear viscosity with the Helfand-moment method.
This expression differs from the Helfand moment by the mere
exchange of a square and a sum over the particles.
The equivalence of the expression in Ref. \cite{McQuarrie}
with the Helfand moment depends on the vanishing of some
cross terms as pointed out in Ref. \cite{CD}.  
Numerical evidence has been obtained in Refs. \cite{A,erpenbeck}
that the expression in Ref. \cite{McQuarrie} 
is in general not valid to calculate the shear viscosity.
We notice that both the original Helfand moment
and the expression in Ref. \cite{McQuarrie} do not
strictly apply to systems subject to p. b. c. 
(see discussions in Refs. \cite{A,erpenbeck}).
This problem is solved by the expression
(\ref{alder-expression}) of Ref. \cite{alder}
in the case of hard-ball fluids and by our expression
(\ref{helfand-torus}) in the general case.

%%%%%%%%%%%%%%%%%%%%%%%%%%%%%%%%%%%%%%%%%%%%%%%%%%%%%%%%%%%%%%%%%%%%%%%%%%%%%%%%

\subsection{Symmetry considerations in two-dimensional systems}

By symmetry, most of the elements of the viscosity tensor are either equal
or vanish.  First, we have:
\begin{equation}
\eta_{ij,kl}=\eta_{kl,ij}=\eta_{ji,kl}=\eta_{ij,lk} \; ,
\end{equation}
because of the stationarity of the equilibrium average,
the reversibility of the microscopic equations, and the fact that
${\bf F}({\bf r}_{a} - {\bf r}_{b})={\bf F}(\Vert{\bf r}_{a} - {\bf
r}_{b}\Vert)$ is a central force.  Secondly, in our work, the fluid is
invariant under rotations by $\varphi=\frac{\pi}{3}$
for the hexagonal geometry and by $\varphi=\frac{\pi}{2}$ for the square one.
If we define the viscosity tensor as a linear operator $\hat\eta$ acting on
matrices $\mathsf A$ according to $({\hat\eta} {\mathsf A})_{ij}=\eta_{ij,kl}
\; A_{kl}$. Then our discrete symmetry can be written as:

\begin{equation}
{\hat \eta}({\mathsf R}^{-1}{\mathsf AR})={\mathsf
R}^{-1}({\hat\eta}{\mathsf A}){\mathsf R} \; ,
\end{equation}
for all matrices $\mathsf{A}$, $\mathsf{R}$ being the rotation matrix

\begin{equation}
{\mathsf R}=
\pmatrix{
\cos \varphi & -\sin \varphi \cr
\sin \varphi & \cos \varphi} \; ,
\label{rotation-matrix}
\end{equation}
and $\varphi$ is equal to $\frac{\pi}{3}$ or $\frac{\pi}{2}$ respectively
for the hexagonal or square systems. Thanks to this symmetry, the only
nonvanishing elements are $\eta_{ij, ij}=\eta_{ji, ij}$
and $\eta_{ii, ii}= \eta_{jj, jj}$.
Furthermore, for $i \neq j$, $k \neq l$,

\begin{equation}
\eta_{ij, ij}
=\eta_{kl,kl }\; , \qquad
\eta_{ii, ii}
=\eta_{jj, jj}\; , \qquad
\eta_{ii, jj}
=\eta_{kk,ll}\; .
\end{equation}
Hence, there are in fact only three independent elements: $\eta_{xx,xx}$,
$\eta_{xy,xy}$, $\eta_{xx,yy}$. On the other hand, for an isotropic system, we
can see that:

\begin{eqnarray}
\eta &=& \eta_{xy,xy} \; ,\\
\zeta &=& \frac{1}{2} \; \left (\eta_{xx,xx} + \eta_{xx,yy} \right ) \; .
\end{eqnarray}
The third element $\eta_{xx,yy}$ is in fact a combination of the two other
elements:

\begin{equation}
\eta_{xx,yy}=\eta_{xx,xx}-2 \; \eta_{xy,xy} \; .
\label{relation-isotropique}
\end{equation}

%%%%%%%%%%%%%%%%%%%%%%%%%%%%%%%%%%%%%%%%%%%%%%%%%%%%%%%%%%%%%%%%%%%%%%%%%%%%%%%%
%%%%%%%%%%%%%%%%%%%%%%%%%%%%%%%%%%%%%%%%%%%%%%%%%%%%%%%%%%%%%%%%%%%%%%%%%%%%%%%%

\section{Description of the two-hard-disk model}
\label{sec.2disks}

\begin{figure}[ht]
\centerline{\epsfig{file=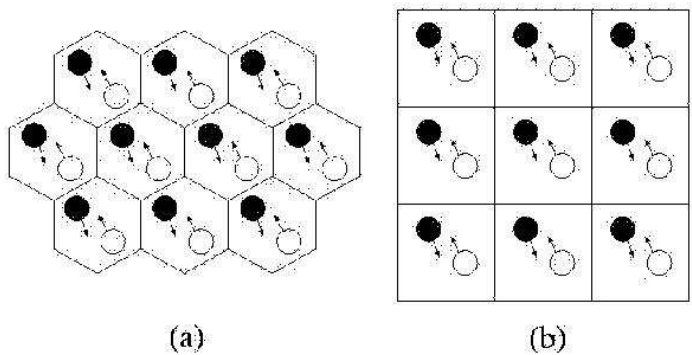,width=10cm}}
\vspace*{0.3cm}
\caption{The model of two hard disks: (a) in the hexagonal geometry and (b)
in the square geometry.}
\label{model-buni-hex-carre}
\end{figure}

In the present work, we apply our method to a simple model which we
describe in the present section.  The model is composed of two hard disks in
elastic collisions on a torus. Bunimovich and Spohn have previously studied
this model for a square geometry \cite{buni-spohn}.  By periodicity,
the system extends to a two-dimensional lattice made of infinitely many
images of the two disks.  For p. b. c. on a square domain, the infinite images
form a square lattice, in which each cell contains two disks (see Fig.
\ref{model-buni-hex-carre}b).

In the present work, we generalize this model to the hexagonal geometry (see
Fig.\ref{model-buni-hex-carre}a). The possibility of such a model was pointed out
in Ref. \cite{gasp-book}.  The images of each disk form now a triangular lattice. 
The two disks (the white and the black ones) have the same diameter $\sigma$ and
mass $m$.  They follow different trajectories. All the black disks move together and
all the white ones move together.  The system is periodic and the dynamics of the
disks can be reduced to the dynamics in the unit cell or torus.

\subsection{Hexagonal geometry}

Let us first introduce some parameters of the system. $L$ is the distance
between the centers of two neighboring cells. It also corresponds to the
distance between two opposite boundaries of a cell.

\begin{figure}[ht]
\centerline{\epsfig{file=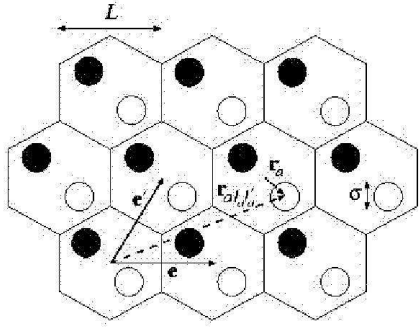,width=6cm}}
\vspace*{0.3cm}
\caption{Basis vector (${\bf e}$ and ${\bf e}'$), position vector ${\bf r}_{a}$ 
of particle $a$ in the cell and the position vector 
${\bf r}_{a \: l_{a} \: l_{a}'}$ in the lattice.}
\label{4.4}
\end{figure}

By a linear combination of two vectors:

\begin{equation}
\cases{
{\bf e}=(L,0) \; ,\cr
{\bf e}'=\left(\frac{1}{2}L,\frac{\sqrt{3}}{2}L\right) \; ,}
\label{basis-vectors}
\end{equation}
we can spot all the cells of the lattice and then localize the center of a
disk thanks to:
\begin{equation}
{\bf r}_{a \; l_{a} \; l_{a}'}={\bf r}_{a} +  l_{a}\; {\bf e}  +  l_{a}'\;
{\bf e}' \; , \quad {\rm for} \ \ a=1,2 \; ,
\end{equation}
where $l_{a}$ and $l_{a}'$ are integer, and ${\bf r}_{a}$ is the position
vector of the disk $a$ with respect to the center of the cell. Therefore, the
distance between the two disks is expressed by

\begin{equation}
\Vert \: {\bf r}_{1 \: l_{1} \: l_{1}'} - {\bf r}_{2 \: l_{2} \: l_{2}'}\:
\Vert =
\Vert \: \underbrace{{\bf r}_{1} -{\bf r}_{2}}_{\bf r} + \\
(l_{1}-l_{2}) \: {\bf e} + (l_{1}'-l_{2}') \: {\bf e}' \: \Vert \; ,
\label{dist-disques}
\end{equation}
where ${\bf r} = {\bf r}_{1} -{\bf r}_{2}$ is the relative position between
both disks.
By the minimum image convention, the relative distance $\Vert{\bf r}\Vert$
should take the smallest value among the infinitely many possible values. Of
course, this distance has to be greater than  or equal to the disk diameter
($\Vert{\bf r}\Vert=\Vert{\bf r}_{1} -{\bf r}_{2}\Vert \geq \sigma$). As we
have a hard-disk potential, the disks move in a free motion between each
collision. Therefore, the equations of motion are written as:

\begin{equation}
\cases{\displaystyle
{{d{\bf r}_{1}}\over{dt}}={{\bf p}_{1}\over m} +\sum_s \Delta{\bf
r}_{1}^{(s)} \; \delta(t-t_s) \; ,\cr
\displaystyle
{{d{\bf r}_{2}}\over{dt}}={{\bf p}_{2}\over m} +\sum_s \Delta{\bf
r}_{2}^{(s)} \; \delta(t-t_s)\; ,\cr}
\label{Newton-position}
\end{equation}

\begin{equation}
\cases{\displaystyle
\frac{d{\bf p}_{1}}{dt}={\bf F}_{1} \; ,\cr
\displaystyle
\frac{d{\bf p}_{2}}{dt}={\bf F}_{2} \; ,\cr}
\label{Newton-momentum}
\end{equation}
where ${\bf p}_{1}$ and ${\bf p}_{2}$ are the momenta of the two disks,
${\bf F}_{1}$
and ${\bf F}_{2}$ being the forces applied respectively to the disks $1$
and $2$. These forces
equal zero when $\Vert{\bf r}_{1}-{\bf r}_{2}\Vert>\sigma$ and are
infinitely repulsive when
$\Vert{\bf r}_{1}-{\bf r}_{2}\Vert=\sigma$.  $t_s$ denotes the time of the
jump to satisfy the minimum
image convention.

At this stage, we can do the following change of variables:
\begin{equation}
\cases{
{\bf r}={\bf r}_{1}-{\bf r}_{2} \; ,\cr
\displaystyle
{\bf R}=\frac{{\bf r}_{1}+{\bf r}_{2}}{2} \; ,\cr}
\label{chg-var-1}
\end{equation}

\begin{equation}
\cases{
\displaystyle
{\bf p}=\frac{{\bf p}_{1}-{\bf p}_{2}}{2} \; ,\cr
{\bf P}={\bf p}_{1}+{\bf p}_{2} \: .\cr}
\label{chg-var-2}
\end{equation}
If we introduce the reduced mass $\mu=\frac{m}{2}$, we can write:
\begin{eqnarray}
\mu \; \frac{d{\bf r}}{dt} &=& {\bf p} + \sum_s \mu\; \Delta{\bf r}^{(s)}
\; \delta(t-t_s)=\mu \; {\bf
v} + \sum_s \mu\; \Delta{\bf r}^{(s)} \; \delta(t-t_s) \; , \\
\frac{d{\bf p}}{dt} &=& {\bf F} = {\bf F}_1 = - {\bf F}_2 \; ,
\end{eqnarray}
where ${\bf v}$ is the relative velocity and $\Delta{\bf
r}^{(s)}=\Delta{\bf r}_1^{(s)}-\Delta{\bf
r}_2^{(s)}$. Here we suppose that we are in the reference frame of the mass
center (that is ${\bf
P}=0$). Accordingly, the energy of the system is reduced to:

\begin{equation}
E=\frac{{\bf p}^{2}}{2 \, \mu} \; .
\end{equation}

\begin{figure}[h]
\centerline{\epsfig{file=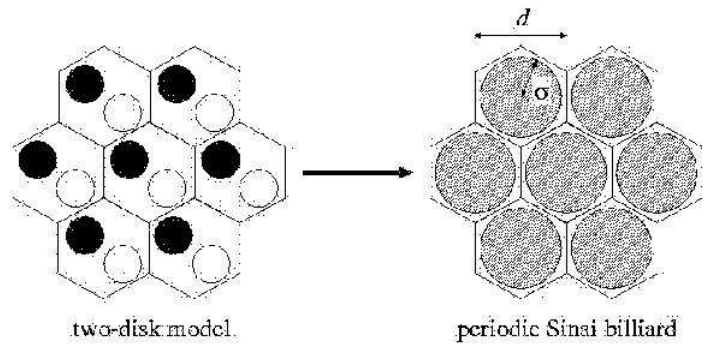,width=10cm}}
\vspace*{0.3cm}
\caption{The model of two hard disks in the hexagonal geometry is reduced to
the periodic Sinai billiard thanks to a change of variables.}
\label{model-change-var-hex}
\end{figure}

The interest of this change of variables is to reduce the number
of variables. Indeed, the only variables that remain are the relative
position and velocity
[${\bf r}=(x,y)$ and ${\bf v}=(v_{x},v_{y})$].
We can associate a fictitious pointlike particle with these variables,
which moves in a reduced
system, known as the \textit{periodic Sinai billiard} (see Fig.
\ref{model-change-var-hex}).

The billiard is also a triangular lattice of hexagonal cells. The size $d$
of these cells is equal to the size of the cells of the model itself ($d=L$). A
hard disk is fixed on the center of each cell. Its radius is equal to the
diameter $\sigma$ of the two moving disks.

The basis vectors of this lattice are the same as those of the original
dynamics (\ref{Newton-position})-(\ref{Newton-momentum}) if we replace $L$ by
$d$, which gives us the possibility to spot a cell in the lattice thanks to the
vector:
\begin{equation}
{\bf r}_{c}=l_{c}\; {\bf e}+l_{c}'\; {\bf e}'\; ,
\end{equation}
where $l_{c}$ and $l_{c}'$ are integer.

In the Sinai billiard, the system is described by a trajectory in a
four-dimensional phase space which are the Cartesian coordinates
($x,y,p_{x},p_{y}$), or the polar coordinates ($x,y,p_{\theta},\theta$).
However, since the energy of the system is conserved, this space is
reduced to the three-dimensional space of the variables ($x,y,\theta$).
Furthermore, in hard-ball systems, the topology of the trajectory is
independent of the energy level. Therefore, we can study the system on an
arbitrary energy level. This energy determines the temperature of the system
and is equal to $E=(d/2)(N-1)k_{\rm B}T=k_{\rm B}T$ because we have only two
degrees of freedom ($d=2$, $N=2$). Sinai and Bunimovich have demonstrated that 
the dynamics in such billiards is ergodic on each energy level
\cite{sinai-buni1,sinai-buni2}.

\subsection{Square geometry}

The case of the square geometry is similar to the
hexagonal one except that the basis vectors are here given by

\begin{equation}
\cases{
{\bf e}=(L,0) \; ,\cr
{\bf e}'=(0,L) \; ,\cr
}
\end{equation}
where $L$ is the length of a side of the square unit cell which contains
two moving disks of diameter
$\sigma$.  We perform the same change of variables to reduce the dynamics
of two hard disks to the one
of the fictitious pointlike particle of a Sinai billiard in a square unit
cell.  Here also, the size
$d$ of the cells of the Sinai billiard is the same as for the cells of the
two hard disks model: $d=L$.

\begin{figure}[h]
\centerline{\epsfig{file=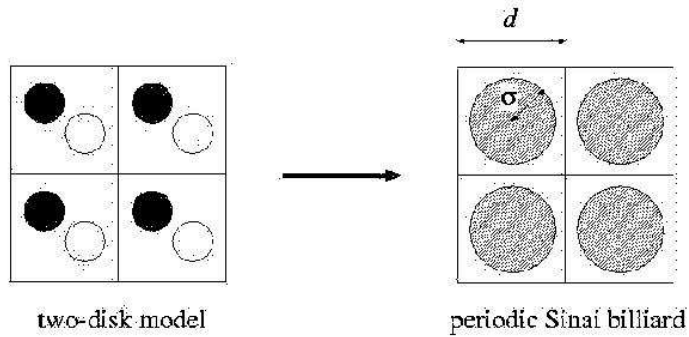,width=10cm}}
\vspace*{0.3cm}
\caption{The model of two hard disks in the square geometry is reduced to the
periodic Sinai billiard thanks to a change of variables.}
\label{model-change-var-carre}
\end{figure}

%%%%%%%%%%%%%%%%%%%%%%%%%%%%%%%%%%%%%%%%%%%%%%%%%%%%%%%%%%%%%%%%%%%%%%%%%%%%%%%%

\subsection{The different dynamical regimes of the model}

The physical quantity determining the size of the cell in our model is the
density  which corresponds to the number of disks per unit volume or, in our
case, the number of disks per unit area. Each cell contains two disks.
Therefore, the density is $n=\frac{2}{V}$ where $V=\Vert{\bf e}\times{\bf
e}'\Vert$ is the area of a cell. In our study, we have chosen that the
diameter of the moving disks is equal to the unity: $\sigma=1$.

\begin{figure}[h]
\centerline{\epsfig{file=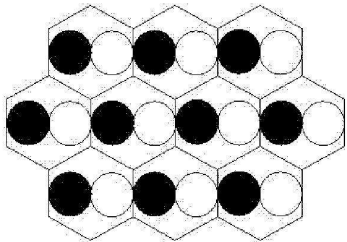,width=5cm}}
\vspace*{0.3cm}
\caption{Hexagonal system at the critical density $n_{\rm cr}$.}
\label{disques-densite-critique}
\end{figure}

As a function of the density, we observe different dynamical regimes. At low density,
the disks are able to move in the whole lattice so that the disks are not localized
in bounded phase-space regions.  In this case, the billiard may have
a finite or an infinite horizon depending on the geometry and on the density.
In the opposite, at high density, the disks are so close to each other that they
cannot travel across the system and we refer to this regime as the {\it localized
regime}. The critical density between the nonlocalized and localized
regimes corresponds to the situation where both disks have a double contact with each
other in the configuration shown in Fig. \ref{disques-densite-critique}.

\subsubsection{Hexagonal geometry}

In the hexagonal geometry, the area of the system is $V=\Vert{\bf e}\times{\bf
e}'\Vert=\frac{\sqrt{3}}{2}L^2$ and the critical density is equal to:
\be
n_{\rm cr}=\frac{\sqrt{3}}{3} \simeq 0.5774 \; ,
\ee
even though the maximum density (the \textit{close-packing density}) is:
\be
n_{\rm max}=\frac{4 \sqrt{3}}{9} \simeq 0.7698 \; .
\ee
At the close-packing density, the system forms a triangular crystal.

In the Sinai billiard, it is well known that there exists different kinds
of regimes according to the dynamics of the particles. As a function of the
density $n$, we observe three regimes:

\begin{enumerate}

\item[$\bullet$]{\textbf{The infinite-horizon regime:} At the low densities
$0 < n < \frac{\sqrt{3}}{4}$, the particles can move in free flight over
arbitrarily large distances. In this regime, the self-diffusion coefficient is
infinite. (See Fig. \ref{mod-lor-hex-inf}.)}

\begin{figure}[h]
\centerline{\epsfig{file=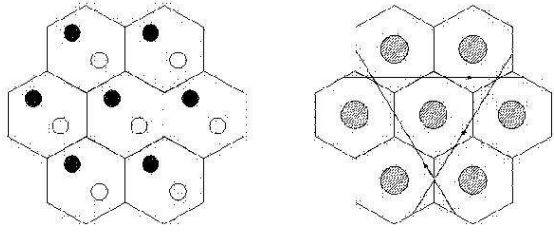,width=8cm}}
\vspace*{0.3cm}
\caption{Typical configuration of the system in the infinite-horizon regime.}
\label{mod-lor-hex-inf}
\end{figure}

\item[$\bullet$]{\textbf{The finite-horizon regime:}
For the intermediate densities $\frac{ \sqrt{3}}{4} < n < n_{\rm cr}$, the
free flights between the collisions are always bounded by a finite distance of
the order of the interdisk distance $d$. Therefore, the horizon is finite and
the self-diffusion coefficient is positive and finite. (See Fig.
\ref{mod-lor-hex-fini}.)}

\begin{figure}[h]
\centerline{\epsfig{file=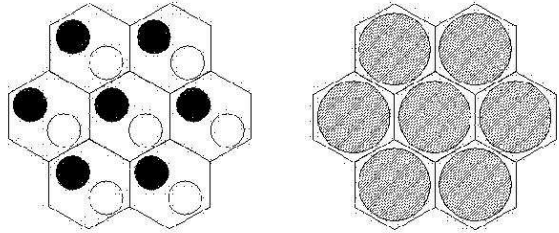,width=8cm}}
\vspace*{0.3cm}
\caption{Typical configuration of the system in the finite-horizon regime.}
\label{mod-lor-hex-fini}
\end{figure}

\item[$\bullet$]{\textbf{The localized regime:} At the highest densities
$n_{\rm cr} < n < n_{\rm max}$, the images of the disk overlap each other in
the billiard so that the relative motion of the particles is localized in
bounded regions. Therefore, the self-diffusion coefficient vanishes. (See
Fig. \ref{mod-lor-hex-local}.)}

\begin{figure}[h]
\centerline{\epsfig{file=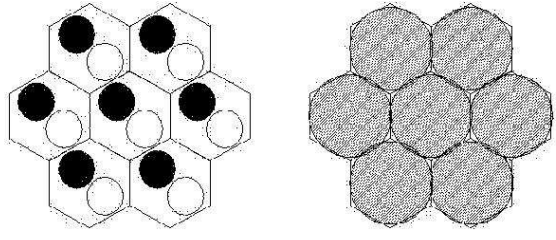,width=8cm}}
\vspace*{0.3cm}
\caption{Typical configuration of the system in the localized regime.}
\label{mod-lor-hex-local}
\end{figure}

\end{enumerate}

We notice that Figs. \ref{mod-lor-hex-inf}, \ref{mod-lor-hex-fini} and
\ref{mod-lor-hex-local} are not depicted at the same scale since the disk
diameter is fixed to unity ($\sigma=1$) and it is the interdisk distance $d$
that varies.

The infinite- and finite-horizon regimes extend over the densities $0<n<n_{\rm cr}$. 
The localized regime corresponds to the densities $n_{\rm cr}<n<n_{\rm max}$.  
The following figure \ref{domaine-hex} shows the different regimes in the
hexagonal geometry.  The remarkable feature of the hexagonal geometry is that there
exists a finite-horizon regime which is not localized, in contrast to the square
geometry (see below).

\begin{figure}[h]
\centerline{\epsfig{file=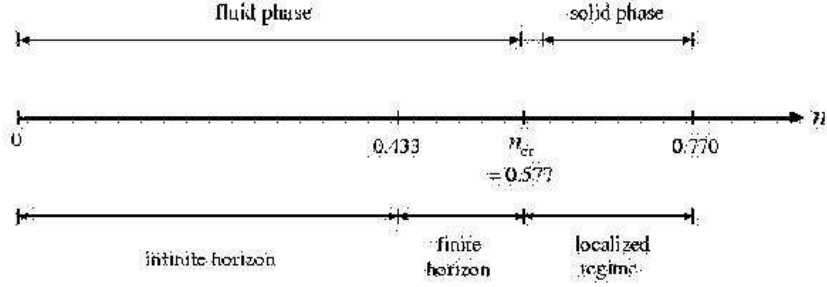,width=12cm}}
\vspace*{0.3cm}
\caption{The different dynamical regimes and thermodynamic phases of the model in
the hexagonal geometry versus the density $n$.}
\label{domaine-hex}
\end{figure}

%%%%%%%%%%%%%%%%%%%%%%%%%%%%%%%%%%%%%%%%%%%%%%%%%%%%%%%%%%%%%%%%%%%%%%%%%%%%%%%%

\subsubsection{Square geometry}

\begin{figure}[h]
\centerline{\epsfig{file=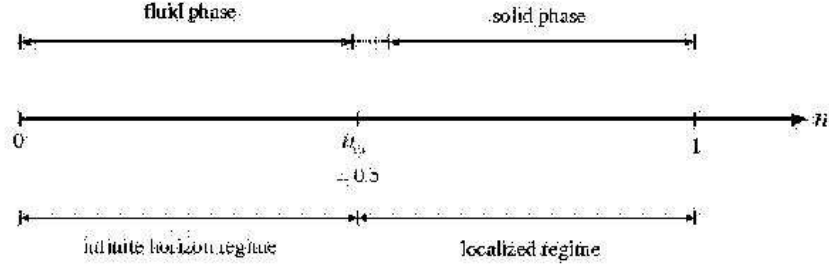,width=12cm}}
\vspace*{0.3cm}
\caption{The different dynamical regimes and thermodynamic phases of the model in
the square geometry versus the density $n$.}
\label{domaine-carre}
\end{figure}

In the square geometry, the volume is $V=\Vert{\bf e}\times{\bf e}'\Vert=L^2$
and the critical density is:
\be
n_{\rm cr}=0.5 \; ,
\ee
which is the density of the transition between the infinite-horizon and the
localized regimes. The close-packing density is equal to:
\be
n_{\rm max}=1\; .
\ee
In Fig. \ref{domaine-carre}, we have depicted the different regimes
in the square geometry.  In the square geometry, there also exist
nonlocalized and localized regimes, but the horizon
is always infinite in the nonlocalized regime. Therefore, it is
only in the localized regime that the horizon is finite in the square geometry.  
This is an important difference with respect to the hexagonal geometry.

%%%%%%%%%%%%%%%%%%%%%%%%%%%%%%%%%%%%%%%%%%%%%%%%%%%%%%%%%%%%%%%%%%%%%%%%%%%%%%%%
%%%%%%%%%%%%%%%%%%%%%%%%%%%%%%%%%%%%%%%%%%%%%%%%%%%%%%%%%%%%%%%%%%%%%%%%%%%%%%%%

\section{Properties of the model}
\label{sec.2disks.prop}

\subsection{Mean free path}

The \textit{mean free path} $\langle l\rangle$ is the
average distance between two successive collisions.  It is known that, in
two-dimensional billiards,
the mean free path is related to the area $\cal A$ of the billiard and its
perimeter $\cal L$ according
to \cite{machta-zwanzig}
\begin{equation}
\langle l \rangle=\frac{\pi {\cal A}}{\cal L} \; . \label{colltime-eq}
\end{equation}
In the different regimes, the mean free path is given by
\begin{enumerate}
\item[$\bullet$]hexagonal geometry:
	\begin{enumerate}
	\item[$(i)$]$\langle l \rangle= \frac{1}{n} - \frac{\pi}{2}\; ,
\qquad n \leq n_{\rm cr}\; ,$
	\item[$(ii)$]$\langle l \rangle= \pi \frac{\frac{2}{n} -\pi +
6\arccos(\frac{1}{\sqrt{\sqrt{3} \; n}})
	- \frac{6}{\sqrt{\sqrt{3} \; n}}\sqrt{1 - \frac{1}{\sqrt{3} \; n}}}
	{2\pi - 12 \arccos(\frac{1}{\sqrt{\sqrt{3}\; n}})} \; ,\qquad n
\geq n_{\rm cr} \; ;$
	\end{enumerate}
\item[$\bullet$]square geometry:
	\begin{enumerate}
	\item[$(i)$]$\langle l \rangle= \frac{1}{n} - \frac{\pi}{2} \; ,
\qquad n \leq n_{\rm cr}\; ,$
	\item[$(ii)$]$\langle l \rangle= \pi \frac{\frac{2}{n} - \pi + 4
\arccos(\frac{1}{\sqrt{2
\;n}}) - 2 \sqrt{\frac{2}{n}} \sqrt{1 - \frac{1}{2 \; n}}}{2 \; \pi - 8 \arccos
(\frac{1}{\sqrt{2 \; n}})} \; ,
	 \qquad n \geq n_{\rm cr} \; .$
	\end{enumerate}
\end{enumerate}

\vspace*{0.5cm}

\begin{figure}[h]
\centerline{\epsfig{file=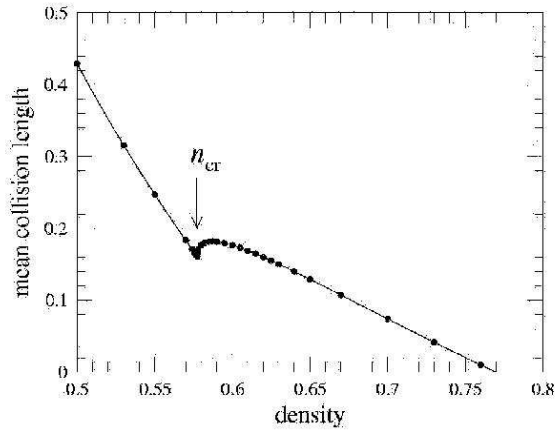,width=8cm}}
\vspace*{0.3cm}
\caption{Theoretical (continuous line) and numerical (dots) values of the
mean free path versus the density $n$ in the hexagonal geometry.}
\label{comparaison-colltime-hex}
\end{figure}

\vspace*{0.5cm}

\begin{figure}[h]
\centerline{\epsfig{file=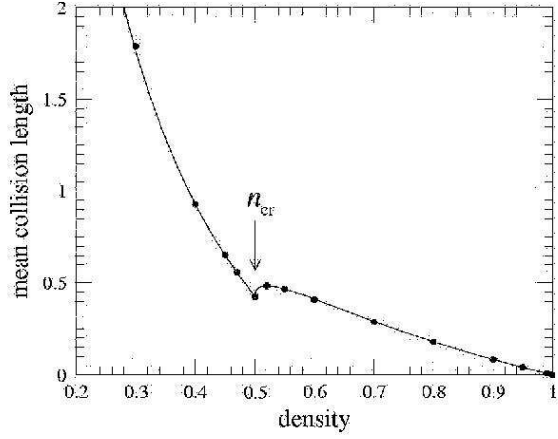,width=8cm}}
\vspace*{0.3cm}
\caption{Theoretical (continuous line) and numerical (dots) values of the
mean free path versus the density $n$ in the square geometry.}
\label{comparaison-colltime-carre}
\end{figure}

We show in Figs. \ref{comparaison-colltime-hex} and
\ref{comparaison-colltime-carre} the excellent
agreement between the above expressions and the values obtained by numerical
simulations.
The break observed in Figs. \ref{comparaison-colltime-hex} and
\ref{comparaison-colltime-carre}
between the nonlocalized and localized regimes can be explained thanks to Eq.
(\ref{colltime-eq}).  Indeed, at the
critical density $n_{\rm cr}$, the disks form a horn.  Above criticality,
the horn becomes a corner
with a finite angle so that the perimeter $\cal L$ decreases very fast.
But, on the other hand, the
area $\cal A$ remains relatively constant. Therefore the ratio
$\frac{\cal A}{\cal L}$ increases with $n$ until this effect disappears.
At higher densities, the mean free path decreases again.

%%%%%%%%%%%%%%%%%%%%%%%%%%%%%%%%%%%%%%%%%%%%%%%%%%%%%%%%%%%%%%%%%%%%%%%%%%%%%%%%

\subsection{Pressure and the different phases of the model}

The hydrostatic pressure allows us to interpret the different regimes
in terms of thermodynamic phases.  The
pressure can be calculated in terms of the time average of the
Helfand moment as shown in Appendix \ref{appendixC}. In the two-disk model with
$N=2$ and $d=2$, the pressure is given by
\be
PV = k_{\rm B}T + R \; ,
\ee
where the rest can be calculated according to Eq. (\ref{rest-coll}) as
\be
R = \frac{\langle \Delta{\bf p}_1^{(c)}\cdot{\bf
r}_{12}^{(c)}\rangle}{4 \; \langle \Delta t_{c-1,c}\rangle} \; ,
\ee
where $\langle \Delta t_{c-1,c}\rangle$ is the mean intercollisional time.
If we denote by $\phi^{(c)}$ the angle between the velocity at collision
and the normal to the disk
of the Sinai billiard, the average in the numerator becomes
\be
\langle \Delta{\bf p}_1^{(c)}\cdot{\bf r}_{12}^{(c)}\rangle =  m \; v \;
\sigma \; \langle\cos\phi^{(c)}\rangle\; ,
\ee
$\sigma$ being the diameter of the disks.  In the case the total momentum
vanishes, the velocity $\bf v$ of the trajectory in the billiard is related to
the relative momentum $\bf p$, the energy, and the temperature by
\be
E=k_{\rm B}T = \frac{{\bf p}^2}{2\mu} =  \frac{{\bf p}^2}{m} =
\frac{\mu{\bf v}^2}{2} =
\frac{m{\bf v}^2}{4} \; ,
\ee
so that ${\bf v} = 2{\bf p}/m$.  At collision, $\sin\phi^{(c)}$ is
uniformly distributed in the interval $[-1,+1]$ so that
\be
\langle\cos\phi^{(c)}\rangle = \frac{\pi}{4} \; .
\ee
On the other hand, the mean intercollisional time of the billiard is
related to the mean free path $\langle l\rangle$ and the speed 
$v=\Vert{\bf v}\Vert$ by
\be
\langle \Delta t_{c-1,c}\rangle = \frac{\langle l\rangle}{v} \; .
\ee
Gathering the results, we obtain the rest as
\be
R = \frac{\pi \, \sigma \, m \, v^2}{16 \,
\langle l\rangle} = \frac{\pi \, \sigma}{4 \,
\langle l\rangle} \; k_{\rm B} T \; .
\ee
Accordingly, the hydrostatic pressure of the model is given by
\be
PV = k_{\rm B} T \left( 1 + \frac{\pi\sigma}{4\langle l\rangle}\right) =
k_{\rm B} T \left( 1 + \frac{\sigma{\cal L}}{4{\cal A}}\right) \; .
\ee
In our work, we introduce the reduced pressure defined as
\begin{equation}
P^* \equiv \; \beta P \; \frac{V}{N} = \frac{PV}{(N-1)k_{\rm B} T}=  1 +
\frac{\pi\sigma}{4\langle l\rangle} =  1 + \frac{\sigma{\cal L}}{4{\cal
A}} \; .
\end{equation}

\begin{figure}[h]
\centerline{\epsfig{file=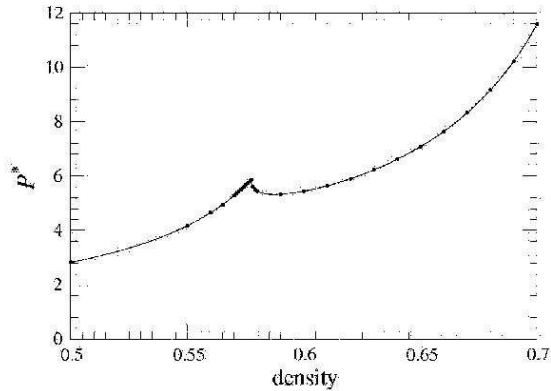,width=8cm}}
\vspace*{0.3cm}
\caption{Theoretical (continuous line) and numerical (dots) values of the
reduced pressure $P^{*}$ versus the density $n$ in the hexagonal geometry.}
\label{pression-hex}
\end{figure}

\begin{figure}[h]
\centerline{\epsfig{file=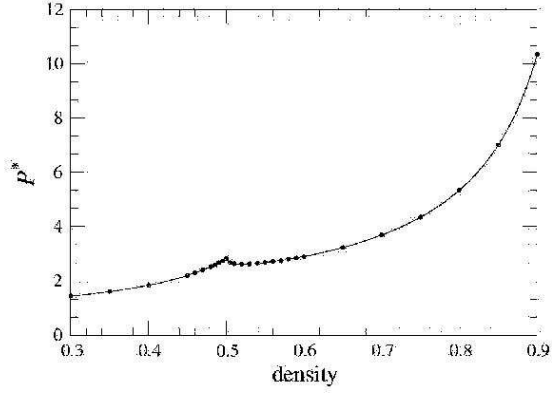,width=8cm}}
\vspace*{0.3cm}
\caption{Theoretical (continuous line) and numerical (dots) values of the
reduced pressure $P^{*}$ versus the density $n$ in the square geometry.}
\label{pression-carre}
\end{figure}

In Figs. \ref{pression-hex} and \ref{pression-carre}, the reduced pressure
is depicted as a function of the density and we observe the manifestation of
a phase transition around the critical density.  The hard-ball systems are
known to present a fluid-solid phase transition 
that we here already observe in the two-disk model.

At low density, the fictitious particle of the Sinai billiard can diffuse in the
whole lattice. This means that the two disks move over arbitrarily large distances
one with respect to the other, which is a feature of a fluid phase.  In contrast,
at high density, the fictitious particle is trapped between three (or four) disks and
its motion is reminiscent of the vibration of atoms in a solid. Of course, it is not
really a vibration since the disks bounce in a chaotic motion because of the elastic
collisions whereas, in a solid, the atoms have quasi-harmonic oscillations around
their equilibrium position.  Nevertheless, we are in the presence of a solid phase
because the translational invariance is broken. Indeed, the motion is no longer
ergodic because the motion now is confined into one among several phase-space
domains of the energy shell.

A phase transition occurs between the fluid and solid phases.
At the critical density $n_{\rm cr}$, the pressure has a
maximum. Above $n_{\rm cr}$, the pressure decreases, reaches a minimum
at a value $n_{\rm cr}'>n_{\rm cr}$, before increasing again.
For $n_{\rm cr}<n<n_{\rm cr}'$, the compressibility would be negative
so that this state would be unstable from a thermodynamic viewpoint.
This suggests a Maxwell construction to determine a fluid-solid
coexistence in the interval of densities $n_{\rm F}<n<n_{\rm S}$
with $n_{\rm F}<n_{\rm cr}$ and $n_{\rm cr}'<n_{\rm S}$.
The values which would delimit this small coexistence interval
in a thermodynamic interpretation of the transition would be
given by
\begin{eqnarray}
\bullet \; \mbox{hexagonal geometry:} \qquad  n_{\rm F} &=& 0.57\pm 0.01 \;
,\nonumber\\ n_{\rm S} &=& 0.60\pm 0.01 \; ,
\end{eqnarray}
and
\begin{eqnarray}
\bullet \; \mbox{square geometry:} \qquad  n_{\rm F} &=& 0.49\pm 0.01 \; ,\nonumber\\
n_{\rm S} &=& 0.55\pm 0.01 \; ,
\end{eqnarray}
(see Figs. \ref{domaine-hex} and \ref{domaine-carre}).
In the square geometry, the horizon is infinite in the fluid phase.
In the hexagonal geometry, the horizon may also be finite
in the fluid phase, which leads to finite viscosity coefficients
in the fluid phase of this model as shown in the following.

%%%%%%%%%%%%%%%%%%%%%%%%%%%%%%%%%%%%%%%%%%%%%%%%%%%%%%%%%%%%%%%%%%%%%%%%%%%%%%%%
%%%%%%%%%%%%%%%%%%%%%%%%%%%%%%%%%%%%%%%%%%%%%%%%%%%%%%%%%%%%%%%%%%%%%%%%%%%%%%%%

\section{Viscosity in the two-hard-disk model}
\label{sec.2disks.visc}

\subsection{The Helfand moment in the two-hard-disk model}

In our model defined with Eqs. (\ref{chg-var-1}) and (\ref{chg-var-2}) and
with a vanishing total
momentum $\mathbf{P}=\mathbf{0}$, the forces obey ${\bf F}_{1} = - {\bf
F}_{2} = {\bf F}$ and the
microscopic current can be written in relative coordinates as

\begin{equation}
J_{ij} = 2 \: \frac{p_{i}p_{j}}{m} + F_{i} \: r_{j}  \; , \label{flux-adapt}
\end{equation}
where ${\bf r}$ is the smallest distance between the disks 1 and 2.
Following the minimum image
convention the position vector presents discontinuities because of the
passages of the relative
position through a boundary, after which it is reinjected into the cell at
the opposite boundary.
We denote the vectors normal to the boundaries of the unit cell by
     \begin{eqnarray}
    \mathrm{hexagonal \ geometry} :
      \cases{
        {\bf c}_{1} = {\bf a} \; ,\cr
        {\bf c}_{2} =- {\bf a} \; ,\cr
        {\bf c}_{3} = {\bf b} \; ,\cr
        {\bf c}_{4} =- {\bf b} \; ,\cr
        {\bf c}_{5} = {\bf b}-{\bf a} \; ,\cr
        {\bf c}_{6} = {\bf a}-{\bf b} \; ,}
      \label{vect-normaux}
     \end{eqnarray}
and
\begin{eqnarray}
   \mathrm{square \ geometry} :
   \cases{
     {\bf c}_{1} = {\bf a} \; ,\cr
     {\bf c}_{2} =- {\bf a} \; ,\cr
     {\bf c}_{3} = {\bf b} \; ,\cr
     {\bf c}_{4} =- {\bf b} \; .\cr
    }
\end{eqnarray}
In order to satisfy the minimum image convention, the relative position
undergoes jumps by vectors which are the vectors normal to the unit cell so
that $\Delta{\bf r}^{(s)} = - {\bf c}_{\omega_s}$ where $\omega_{s}$ denotes
the label of the boundary crossed by the particle at
the $s$\textsuperscript{th} passage at time $t_s$. In these notations,
Hamilton's equations take the form
\begin{eqnarray}
\cases{
\frac{d{\bf r}}{dt} = \frac{2{\bf p}}{m} - \sum_{s}^{}
 {\bf c}_{\omega _{s}} \: \delta (t-t_{s}) \; ,\cr
\frac{d{\bf p}}{dt} = {\bf F} \; . }
\label{eq-mvt-periodique}
\end{eqnarray}

In this periodic system, the expression for the Helfand moment is given by a
reasoning similar to the one leading to Eq. (\ref{helfand-torus}).  We obtain:

\begin{equation}
G_{ij}(t) =  p_{i}(t) \: r_{j}(t) +
\sum_{s}^{}  p_{i}(t_{s}) \:
{c}_{\omega _{s} j} \: \theta (t-t_{s})
\; . \label{moment-helfand}
\end{equation}
Finally, the viscosity coefficients have the expressions:

\begin{eqnarray}
\eta_{ij,kl}  = \lim_{t \to \ \infty } \frac{\beta}{2tV} \left( \left \langle
\sum_{t_s<t} p_{i}(t_{s}) \: {c}_{\omega _{s} j}
 \sum_{t_{s'}<t} p_{k}(t_{s'}) \: {c}_{\omega _{s'} l} \right \rangle
- \left \langle \sum_{t_s<t} p_{i}(t_{s}) \: {c}_{\omega _{s} j} \right \rangle
\left \langle \sum_{t_{s'}<t} p_{k}(t_{s'}) \: {c}_{\omega _{s'} l} \right
\rangle \right) \; .
\label{visc-helf-per}
\end{eqnarray}
Let us remark that the terms $p_{i}(t) \; r_{j}(t)$ do not appear in this
relation because they do not contribute to the viscosity coefficients. Indeed,
the relative position ${\bf r}(t)$ and momentum ${\bf p}(t)$ remain bounded in
the course of time and their contribution disappears in the limit $t\to\infty$.

In the following, the numerical results are presented in terms of a
reduced viscosity tensor which is defined by
\begin{equation}
\eta_{ij,kl}^{*} \equiv \frac{\eta_{ij,kl}}{2\sqrt{mk_{\rm B}T}} \; .
\end{equation}

\subsection{Hexagonal geometry}

\begin{figure}[h]
\centerline{\epsfig{file=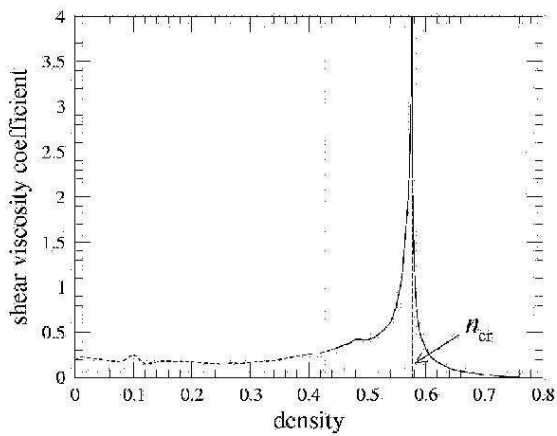,width=8cm}}
\vspace*{0.3cm}
\caption{Shear viscosity coefficient $\eta^*$ versus the density in the hexagonal
geometry. The part in dashed line corresponds to the density in which 
the coefficient would not exist in the limit $t\to\infty$ because the horizon
is infinite. The long dashed vertical lines separate  the different regimes: on
the left-hand side, the horizon-infinite regime  (fluid phase); at the center,
the horizon-finite regime (fluid phase); and on the right-hand side, localized
regime (solid phase).}
\label{viscosite-hex}
\end{figure}

\begin{figure}[h]
\centerline{\epsfig{file=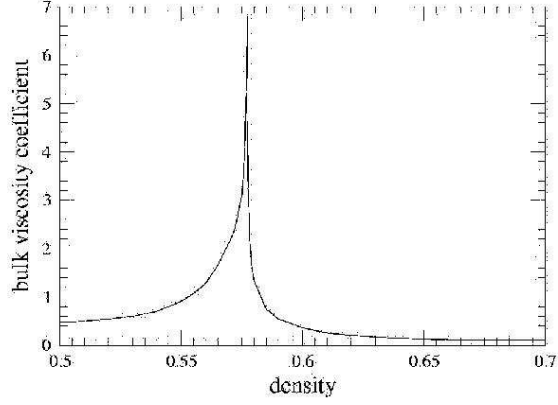,width=8cm}}
\vspace*{0.3cm}
\caption{Bulk viscosity coefficient $\zeta^*$ versus the density in the hexagonal
geometry.}
\label{bulk-viscosite-hex}
\end{figure}

\begin{figure}
\centerline{\epsfig{file=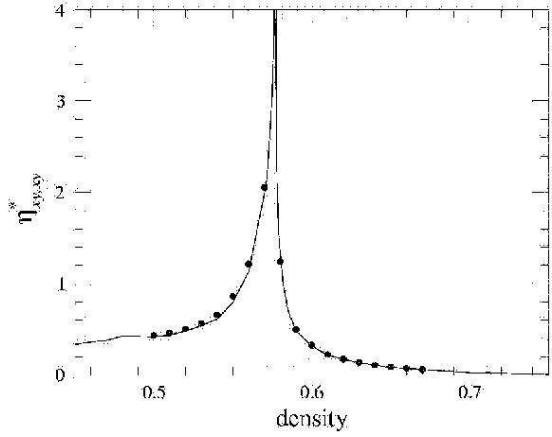,width=8cm}}
\vspace*{0.3cm}
\caption{Tensor element $\eta_{xy,xy}^{*}$ of shear viscosity versus the
density in the hexagonal geometry. The dots represent the results of the
relation (\ref{relation-isotropique}): $\eta_{xy,xy}^{*}=\frac{1}{2} \left
(\eta_{xx,xx}^{*} - \eta_{xx,yy}^{*} \right )$.  The continuous line
corresponds to the data of Fig. \ref{viscosite-hex}.}
\label{relation-isotropiqueG}
\end{figure}

In the hexagonal geometry the fourth-order tensor of viscosity is isotropic.
Indeed, since the system is invariant under rotations by $\frac{\pi}{3}$, we
obtain the relation $\eta_{xx,yy}=\eta_{xx,xx}-2 \; \eta_{xy,xy}$ 
which implies the full rotation invariance of the
viscosity tensor. We depict in
Figs. \ref{viscosite-hex} and \ref{bulk-viscosite-hex} the results obtained
for the reduced viscosities ($\eta^{*}$, $\zeta^{*}$) and the relation
(\ref{relation-isotropique}) is checked in
Fig. \ref{relation-isotropiqueG}.

\vspace*{1cm}

%\vspace*{0.5cm}

In the infinite-horizon regime, the trajectory can present arbitrarily
large displacements in the system without undergoing any collision.
Accordingly, the variance of the Helfand moment $G_{yx}$ increases faster than
linearly as $t \; \log t$, which implies an infinite viscosity coefficient
after averaging over an infinite time interval. However, the factor $\log t$
generates a so weak growth that it does not manifest itself much over
the finite time of the simulation. This is the reason why we obtain finite
values for the viscosity coefficients in Figs. \ref{viscosite-hex} and
\ref{bulk-viscosite-hex}.  However, these values are only indicative since
they should be infinite, strictly speaking.

On the other hand, in the finite-horizon regime, the variance of the
Helfand moment has a strictly linear increase in time and the viscosity
coefficients are finite and positive.  This is the result of a central-limit
theorem which holds in the finite-horizon regime of the hexagonal geometry,
as can be proved by considerations similar to those developed by Bunimovich
and Spohn \cite{buni-spohn}.  We observe in Fig. \ref{viscosite-hex} that
the viscosity has a diverging singularity at the critical density
($n_{\rm cr}=\frac{\sqrt{3}}{3}$) which corresponds to the fluid-solid
phase transition.  We shall explain below the origin of this singularity.

Finally, in the localized regime corresponding to the solid phase, the
viscosity is finite and
positive, and decreases when the density increases until the maximum density.

\vspace*{0.5cm}

\subsection{Square geometry}

\begin{figure}[h]
\centerline{\epsfig{file=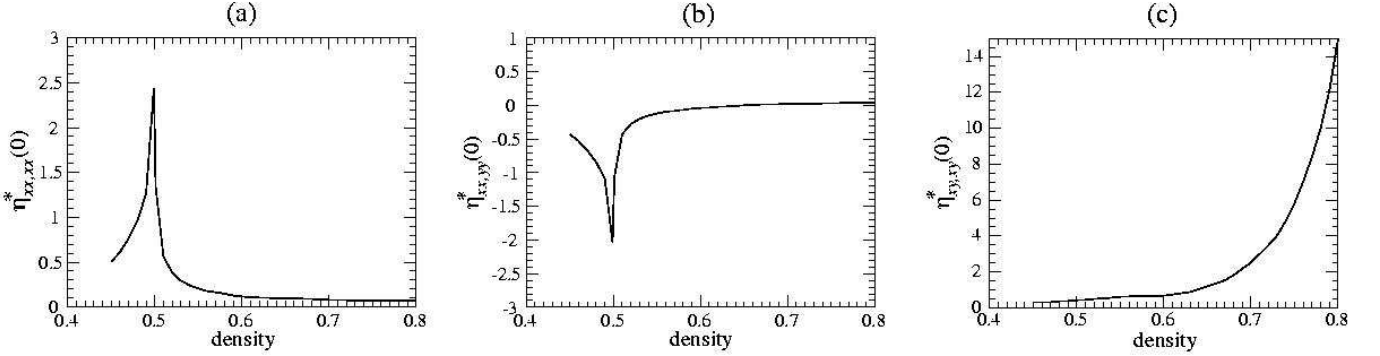,width=18cm}}
\vspace*{0.3cm}
\caption{Square geometry: The three independent tensor elements
(a) $\eta_{xx,xx}^*$, (b) $\eta_{xx,yy}^*$, (c) $\eta_{xy,xy}^*$ for $\varphi=0$.}
\label{visc-carre-0}
\end{figure}

\begin{figure}[h]
\centerline{\epsfig{file=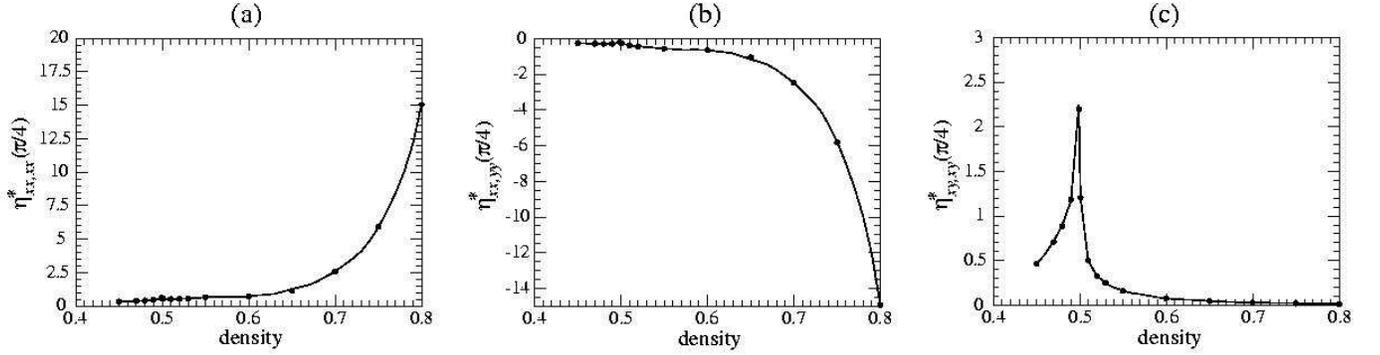,width=18cm}}
\vspace*{0.3cm}
\caption{Square geometry: The three independent tensor elements (a) $\eta_{xx,xx}^*$,
(b) $\eta_{xx,yy}^*$, (c) $\eta_{xy,xy}^*$ for $\varphi=\frac{\pi}{4}$.
The continuous line corresponds to the results obtained numerically and 
the dots to the values obtained by the relations (\ref{combi-45}).}
\label{visc-carre-45}
\end{figure}

In the square geometry, the fourth-order viscosity tensor is not isotropic.
Indeed, the tensor is transformed by the matrix $R_{ij}(\varphi)$ of rotation
by an angle $\varphi$ into

\begin{equation}
\eta_{ij,kl}(\varphi)=R_{i i'}(\varphi) \; R_{j j'}(\varphi) \; R_{k
k'}(\varphi) \; R_{l l'}(\varphi) \; \eta_{i'j',k'l'}(0) \; .
\end{equation}
For example, if $\varphi=\frac{\pi}{4}$, we have:

\begin{equation}
\cases{
\eta_{xx,xx}(\frac{\pi}{4})=\frac{1}{2}
\left[\eta_{xx,xx}(0) + \eta_{xx,yy}(0) + 2 \; \eta_{xy,xy}(0)\right] \; ,\cr
\eta_{xy,xy}(\frac{\pi}{4})=\frac{1}{2} \left[\eta_{xx,xx}(0) -
\eta_{xx,yy}(0) \right] \; ,\cr
\eta_{xx,yy}(\frac{\pi}{4})=\frac{1}{2} \left[\eta_{xx,xx}(0) +
\eta_{xx,yy}(0) \right]
-\eta_{xy,xy}(0) \; . }
\label{combi-45}
\end{equation}
Since the system is not isotropic, one more viscosity coefficient is
required beside the shear and bulk viscosities. Therefore, we have to evaluate
the three independent tensor elements $\eta_{xx,xx}, \eta_{xy,xy},
\eta_{xx,yy}$ which are depicted in Figs. \ref{visc-carre-0} and
\ref{visc-carre-45} with respect to two different axis frames: in
the first one the axes are parallel to the sides of the square
($\varphi=0$) and, in the second one, they form an angle of 45 degrees with
respect to the lattice ($\varphi=\frac{\pi}{4}$).  Figure \ref{visc-carre-45}
shows that the relations (\ref{combi-45}) are well satisfied between the
elements of the viscosity tensor.

In the square geometry, Bunimovich and Spohn have proved a central-limit
theorem for viscosity in the localized regime which coincides with the solid
phase above the critical density \cite{buni-spohn}.  In this range of density,
the viscosity coefficient is thus guaranteed to be positive and finite.

In the fluid phase, the horizon is infinite and the viscosity is infinite
because of a growth as $t\log t$ of the variance of the Helfand moment for a
reason similar as in the hexagonal geometry.  In our numerical simulation over
a finite time interval, the viscosity takes finite values because the
logarithmic growth is very slow.

An important difference with respect to the hexagonal geometry is the
absence of a singularity of the viscosity coefficient $\eta_{xy,xy}^{*}(0)$ at
the phase transition in the square geometry.  However, such a singularity still
appears in the square geometry in the coefficients $\eta_{xx,xx}^*(0)$,
$\eta_{xx,yy}^*(0)$, and $\eta_{xy,xy}^*(\frac{\pi}{4})$.

Moreover, in the solid phase, the coefficient $\eta_{xy,xy}^{*}(0)$
increases with the density, as explained here below.

\subsection{Explanation of the numerical observations}

\subsubsection{Solid phase}

The behavior of the viscosity tensor is clearly different in the two
geometries. In this section, we explain these differences by comparing the
topology of the trajectories in both geometries, since these trajectories form
the basis of the evolution of the Helfand moment.  More precisely, we will
compare the behavior of $\eta_{xy,xy}^{*}$ between the hexagonal and square
geometries for $\varphi=0$. This viscosity coefficient is given by

\begin{equation}
\cases{
\eta_{xy,xy}^{*} = \eta_{yx,yx}^{*} \sim \frac{\left \langle
G_{yx}(t)^{2} \right \rangle}{t}
\; , \qquad \left \langle G_{yx}(t) \right\rangle=0 \; ,\cr
G_{yx} \sim \sum_{s} v_{y}(t_s) \; c_{\omega_{s}x} \;,
}
\label{helfand-moment-explain}
\end{equation}
where $v_{y}(t_s)$ is the $y$-component of the velocity at the time $t_s$ of
the jump.

When the density tends to the closed-packing density, the accessible
domain of the particles tends to a perfect triangle in the hexagonal
geometry. On the other hand, in the square geometry, it tends to a perfect
square. This difference is at the origin of the different behaviors of the
$\eta_{xy,xy}^{*}$ in both lattices.

\begin{figure}[h]
\centerline{\epsfig{file=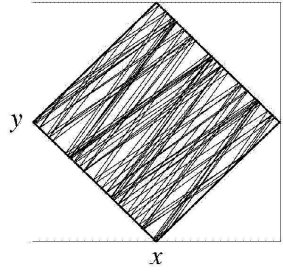,width=4cm}}
\caption{Part of a typical trajectory in the square geometry when the
density tends to the closed-packing density.}
\label{traj-carre-grande-densite}
\end{figure}

First, let us consider the case of the square geometry. In Fig.
\ref{traj-carre-grande-densite}, we depict a typical trajectory of the
fictitious particle moving in the Sinai billiard. We observe
that this trajectory presents a regular motion between two opposite
``walls'' (these walls are made of parts of the fixed hard disks in the
billiard). At the limit where the billiard is a perfect square, the
trajectories will bounce back and forth in a regular motion. Indeed the square
billiard is an \textit{integrable system}.

\begin{figure}[h]
\centerline{\epsfig{file=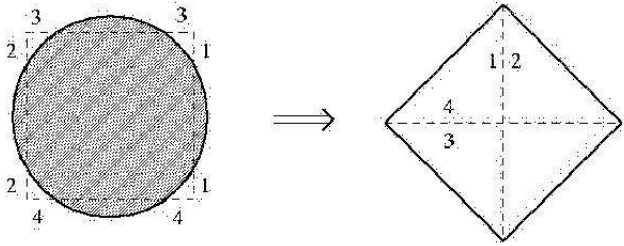,width=9cm}}
\vspace*{0.3cm}
\caption{Geometry and notation for the boundaries 
in the case of the square geometry at high density.}
\label{def-parois-carre}
\end{figure}

As we have seen before, the evolution of the Helfand moment along the
trajectories is determined by the passages through the boundaries (see Fig.
\ref{def-parois-carre}).  Both horizontal boundaries (3 and 4) do not 
contribute to the evolution of $G_{yx}$ since the $x$-component of the normal
vectors to these boundaries equals zero. Therefore, only the passages through
the vertical boundaries contribute  to the Helfand moment in the square
geometry.

\begin{figure}[h]
\centerline{\epsfig{file=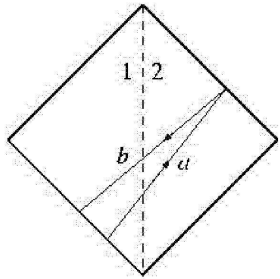,width=4cm}}
\vspace*{0.3cm}
\caption{Part of a typical trajectory in the square geometry at high density.}
\label{portion-traj-carre-grande-densite}
\end{figure}

To understand the behavior of the Helfand moment, let us take a small part
of the typical trajectory drawn in Fig. \ref{traj-carre-grande-densite} (see
Fig. \ref{portion-traj-carre-grande-densite}).  First, let us consider the part
denoted by the letter $a$ in Fig. \ref{portion-traj-carre-grande-densite}. This
one crosses the boundary in the direction $1\to 2$, which means that
$c_{\omega_{s}x}$ is positive (since $c_{1x}=\frac{d}{2}$). On the other hand,
the $y$-component of the velocity, $v_{y}$, is also positive. Therefore, the
contribution of the small part $a$ to the evolution of $G_{yx}$ is positive.

Now, let us take the part of the trajectory denoted $b$ in Fig.
\ref{portion-traj-carre-grande-densite}. In this case, the particle crosses
the boundary in the direction $2\to 1$ and $c_{2x}$ is negative. Since $v_{y}$
is also negative, the product of these two quantities is positive, and so at
each successive crossings of the boundary $1-2$. Consequently, we obtain a sum
of positive terms and the Helfand moment quickly increases along a trajectory
as the one of Fig. \ref{traj-carre-grande-densite}.

\begin{figure}[h]
\centerline{\epsfig{file=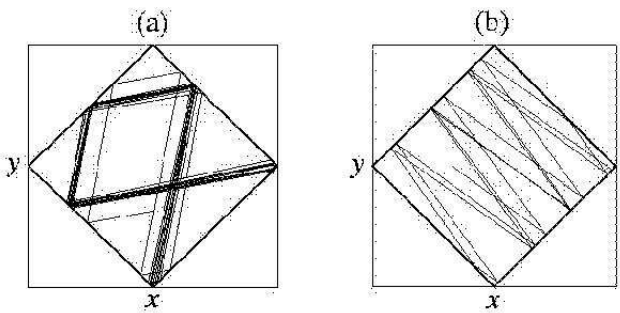,width=9cm}}
\vspace*{0.3cm}
\caption{Square geometry at high density: The trajectory is depicted (a) during a
transient regime before (b) another regime with most bounces on the two other opposite
walls.}
\label{regimes-carre}
\end{figure}

\begin{figure}[h]
\centerline{\epsfig{file=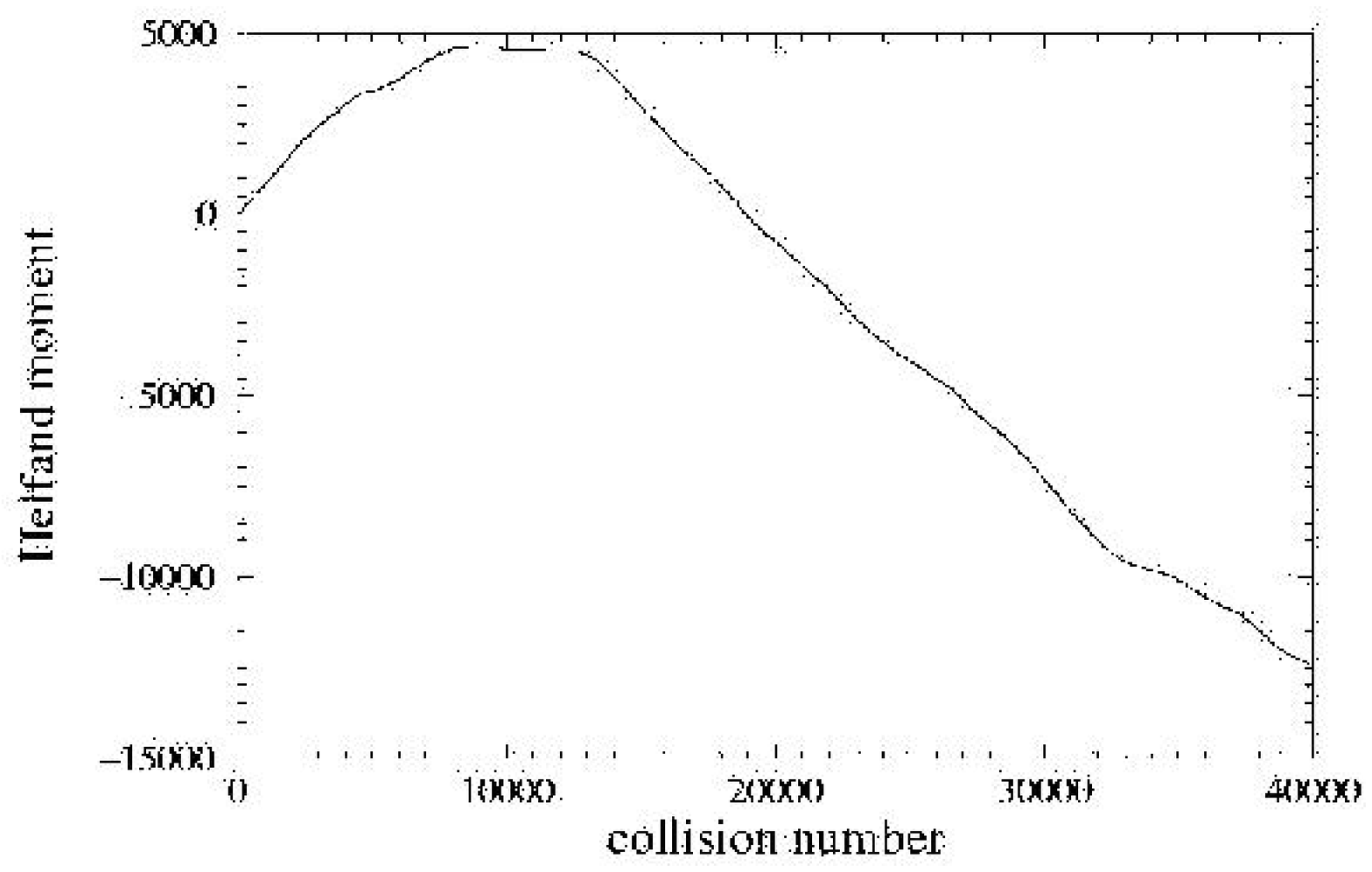,width=10cm}}
\vspace*{0.3cm}
\caption{Evolution of the Helfand moment along a typical trajectory in the
square geometry at high density.}
\label{moment-evolution-carre-grande-densite}
\end{figure}

However, the square is not perfect and the walls are still slightly convex.
Therefore, after a certain time, the trajectory shown in Fig.
\ref{traj-carre-grande-densite} goes into a transient regime shown on the
left-hand side of Fig. \ref{regimes-carre} before another regime in which the
particle collides most often the two other walls (see the right-hand side of
Fig. \ref{regimes-carre}).

With the same reasoning as before, we conclude that the contributions are
negative in this new regime and the Helfand moment decreases during a long time
interval.

The evolution of the Helfand moment along the whole trajectory is depicted
in Fig. \ref{moment-evolution-carre-grande-densite} where we observe the
succession of the three types of regimes which we have described here above.
We notice that the nearly constant part corresponds to the transient regime.

The larger is the density the more perfect is the square and the longer the
trajectory remains in a particular regime. Therefore, the Helfand moment can
have larger and larger variations, which implies an increase of the coefficient
$\eta_{xy,xy}^{*}(0)$ of shear viscosity with density.

\begin{figure}[h]
\centerline{\epsfig{file=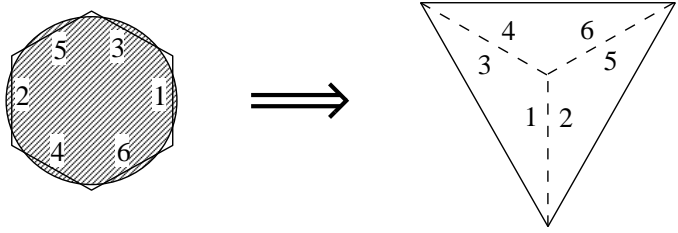,width=9cm}}
\vspace*{0.3cm}
\caption{Geometry and notation for the boundaries in the case of the
hexagonal geometry at high density.}
\label{def-parois-hex}
\end{figure}

\begin{figure}[h]
\centerline{\epsfig{file=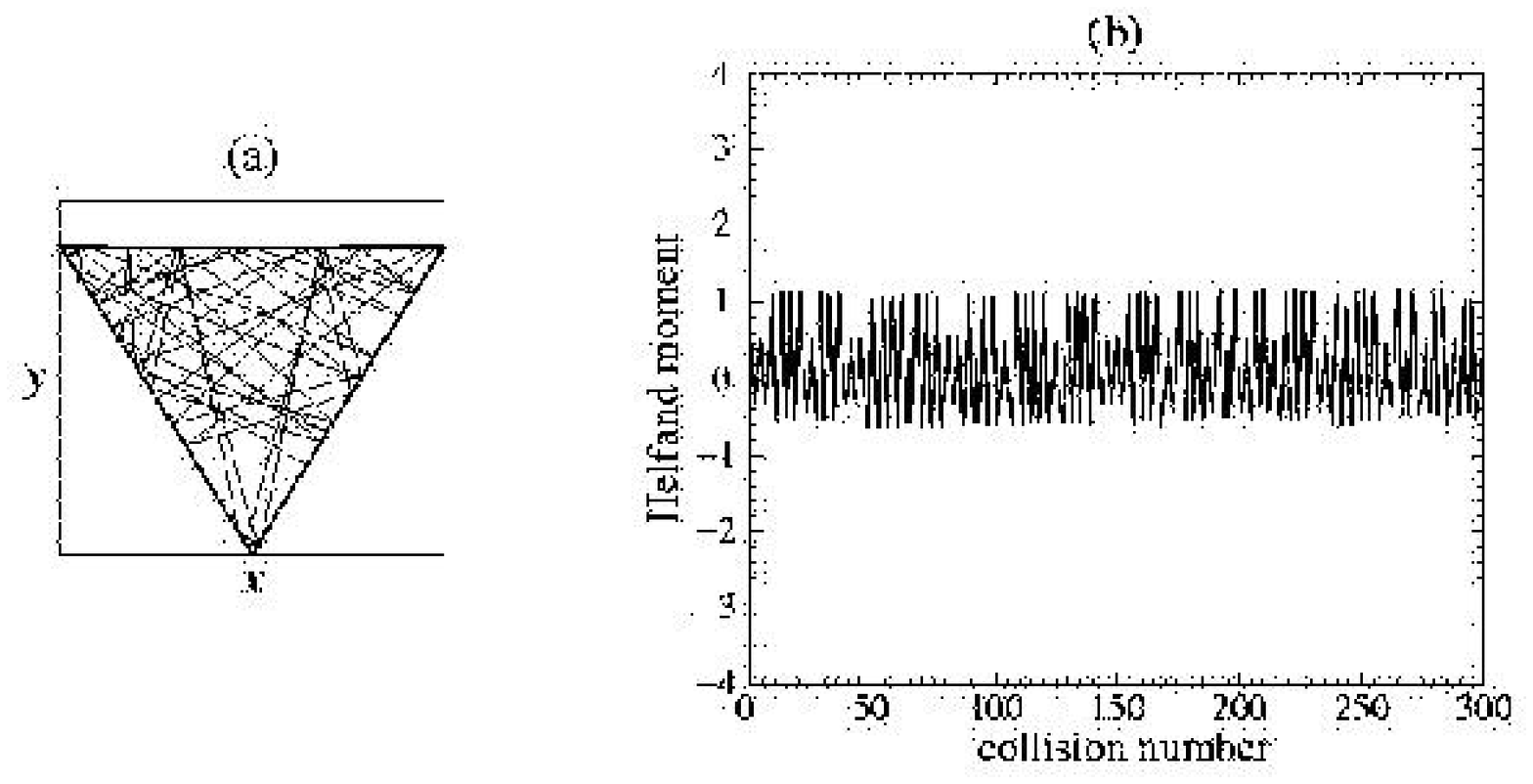,width=12cm}}
\vspace*{0.3cm}
\caption{Hexagonal geometry at high density: (a) Part of
a typical trajectory when the density tends to the closed-packing density. 
(b) Evolution of the Helfand moment along this typical trajectory.}
\label{regimes-hex}
\end{figure}

In the hexagonal geometry (see Fig. \ref{def-parois-hex}), the trajectories
present another behavior. We show in Fig. \ref{def-parois-hex} a typical 
trajectory in this geometry with a density larger than the critical density.
We observe that the trajectory visits the whole billiard in different
directions and therefore goes into very different velocities. 
Accordingly, the particle crosses the boundaries with
random values of its velocity in contrast to its behavior in the square
geometry.  Consequently, the quantity $c_{\omega_{s}x}$ can be positive 
at a particular crossing and negative at the
next one.  Hence the Helfand moment cannot increase or decrease over long
periods as in the square geometry (see Fig. \ref{regimes-hex}). 
This explains qualitatively why, in the solid phase, the
coefficient $\eta_{xy,xy}^{*}(0)=\eta^{*}$ is much smaller in the hexagonal
geometry than in the square one.

In the square geometry with $\varphi=\frac{\pi}{4}$, the same arguments as
in the hexagonal case explain the decrease of
$\eta_{xy,xy}^{*}(\frac{\pi}{4})$ at high density. By the relations between
the different elements of the viscosity tensor, we can also understand the
behavior of the other elements in both geometries.

\vspace*{0.5cm}

\subsubsection{Fluid-solid phase transition}

In both the hexagonal and square geometries, the two-disk model presents a
phase transition.  This transition is reminiscent of the fluid-solid phase
transition in the many-disk system where the viscosity coefficient is also
singular.  In this regard, the two-disk model can contribute to the
understanding of the changes in the transport properties across the
fluid-solid phase transition.

We first explain why $\eta_{xy,xy}^{*}$ presents a diverging singularity at
the critical density in the hexagonal geometry and not in the square geometry
for $\varphi=0$. Here again, we compare the topology of the trajectories in
both geometries and the way in which the Helfand moment evolves along
these trajectories. At densities close to the critical density, both
geometries present what we call \textit{traps}.

\vspace*{0.5cm}

\begin{figure}[h]
\centerline{\epsfig{file=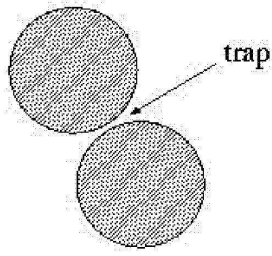,width=4cm}}
\vspace*{0.3cm}
\caption{Example of traps in which the particles can enter and remain a long time.}
\label{pieges-hex-carre-densite-critique}
\end{figure}

\begin{figure}[h]
\centerline{\epsfig{file=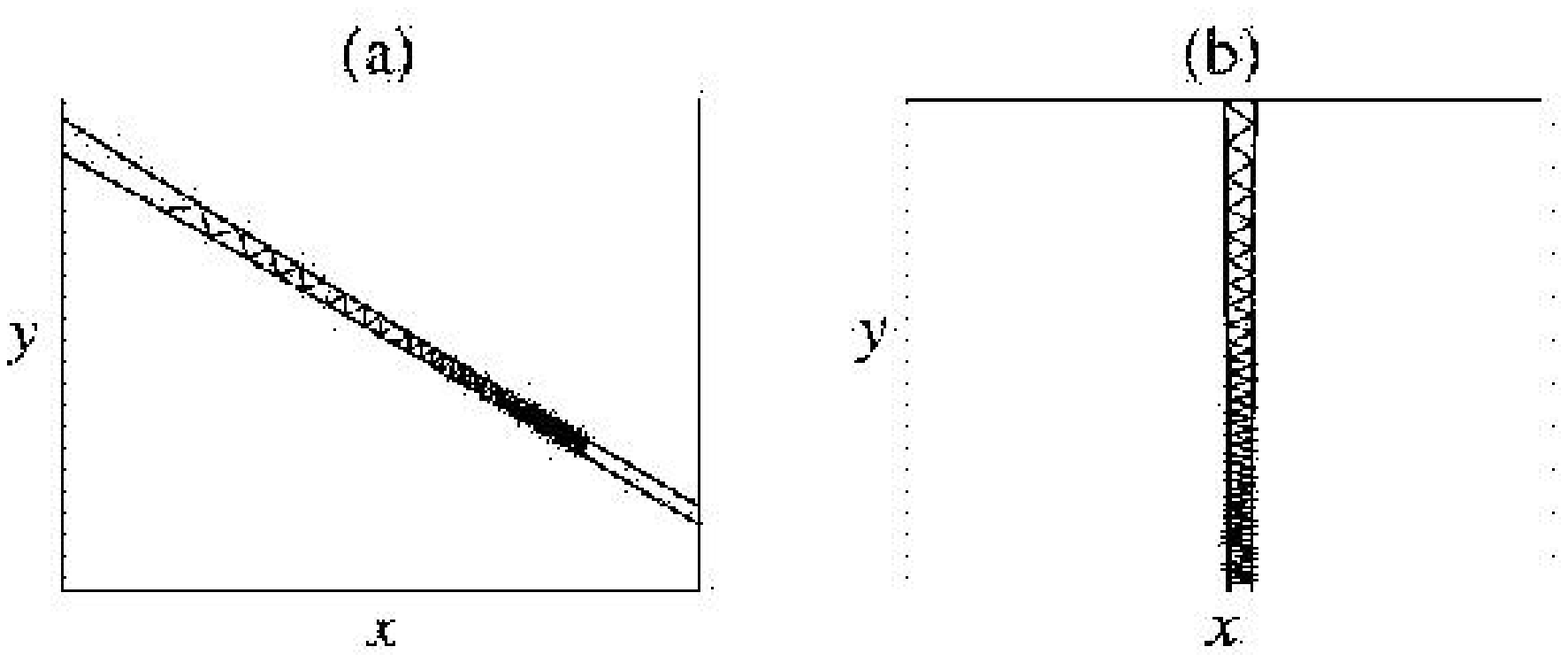,width=10cm}}
\vspace*{0.3cm}
\caption{Particle trapped between two disks very close to each other in the hexagonal
geometry.  The line joining their centers either (a) forms an angle with the horizontal 
or (b) is horizontal.}
\label{particules-piegee-densite-critique}
\end{figure}

Figure \ref{pieges-hex-carre-densite-critique} shows an example of a trap.
These traps are particular regions of the billiard where the particle can
remain during a long time interval. Figure \ref{particules-piegee-densite-critique})
depicts typical examples of a particle moving in such traps.
When the particle travels out of the traps, the Helfand moment does not increase quickly in
both geometries.  Therefore, it is the presence of the traps which is at the origin of the
difference between both geometries.

In the square geometry, the traps do not influence the evolution of
$G_{yx}$. Indeed, as we have already mentioned here above, the passages through
the horizontal boundary $3-4$ do not contribute since $c_{3x} = c_{4x} = 0$ (see Fig.
\ref{def-parois-carre} for the definitions of the boundaries in the square geometry).
Therefore, the horizontal traps around these boundaries do not contribute. There remains the
vertical traps. When a particle bounces for a long time in one of these traps, $c_{1x}$ and
$c_{2x}$ are not vanishing, but the velocity $v_{y}$ is almost equal to zero so that the
vertical traps does not contribute much either. This implies that both kinds of traps
contribute very slightly to the evolution of the Helfand moment. To conclude the Helfand
moment diffuse in the same way as for the other densities and the coefficient
$\eta_{xy,xy}^{*}(0)$ does not present any divergence at the fluid-solid transition in the
square geometry.

\vspace*{0.5cm}

%\vspace*{0.5cm}

\begin{figure}[h]
\centerline{\epsfig{file=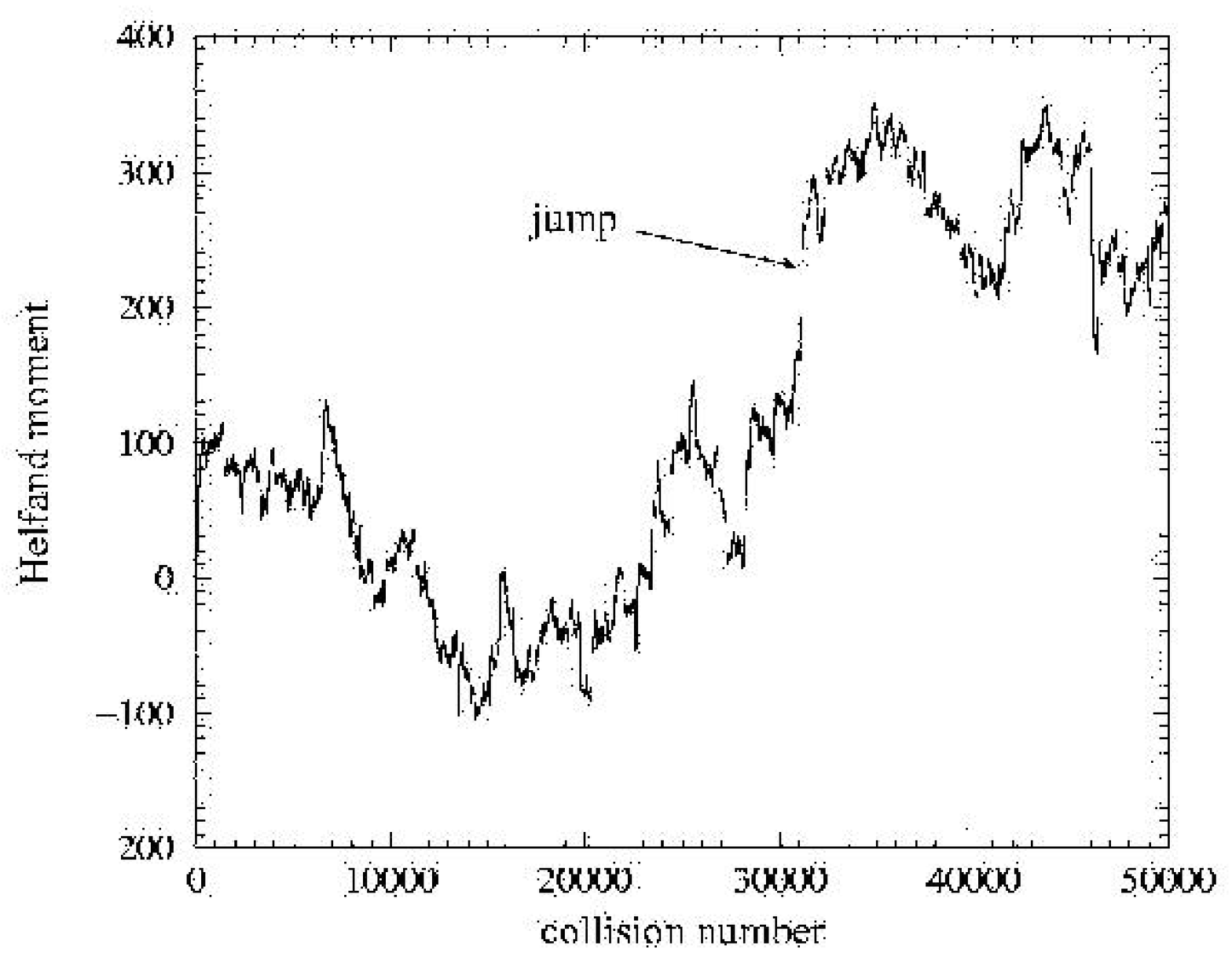,width=9cm}}
\vspace*{0.3cm}
\caption{Helfand moment in the hexagonal geometry evaluated along a
particular trajectory at a density tending to the critical density.}
\label{moment-saut-densite-critique}
\end{figure}

On the other hand, in the hexagonal geometry, the traps along the boundaries making an angle of
$30^{\circ}$ with respect to the horizontal are very important for the evolution of $G_{yx}$,
whereas the vertical traps do not participate significantly. Figure
\ref{moment-saut-densite-critique} shows a typical diffusion of the Helfand moment.  We
observe in Fig. \ref{moment-saut-densite-critique} the presence of jumps which correspond to
the passages in the traps like the one drawn on the left-hand side of Fig.
\ref{particules-piegee-densite-critique}.  Because of these jumps, the Helfand
moment quickly diffuses. Furthermore, the importance of these traps in the
hexagonal geometry can also be understood by comparing the behavior of the Helfand moment as a
function of time at densities below and above the critical one $n_{\rm cr}$.

\begin{figure}
\centerline{\epsfig{file=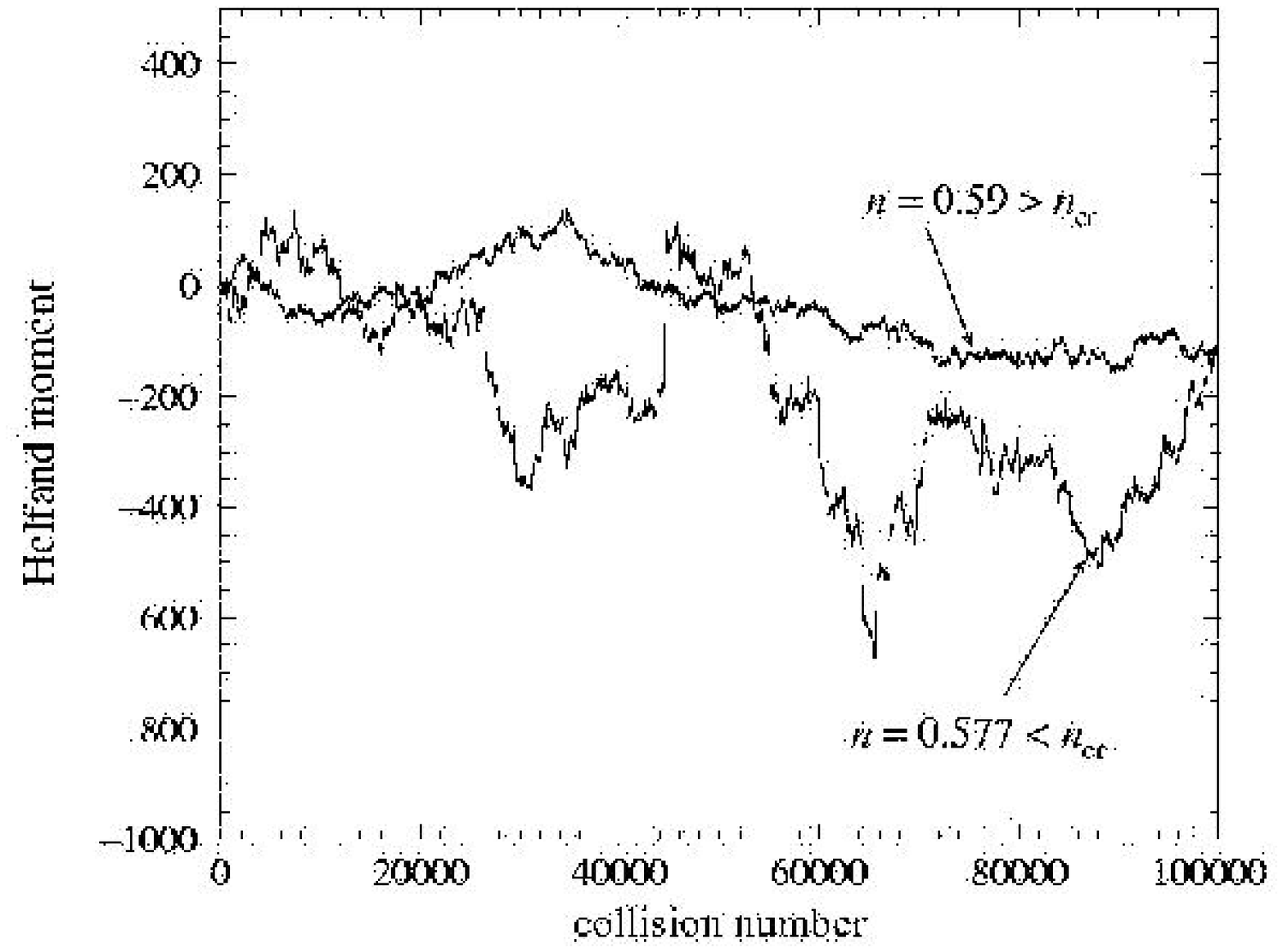,width=9cm}}
\vspace*{0.3cm}
\caption{Comparison of the evolution of the Helfand moment for two
different densities separated by the critical density in the hexagonal geometry.}
\label{comparaison-moment-hex-autour-densite-critique}
\end{figure}

We illustrate this point in Fig. \ref{comparaison-moment-hex-autour-densite-critique}
where we observe that there are no more jumps above the critical density.  Therefore,
$G_{yx}$ does not vary much contrary to the case of densities just below $n_{\rm cr}$.
Above criticality, the size of the traps decreases so quickly that the contribution
of these traps decreases and, thus, the viscosity coefficient $\eta_{xy,xy}^{*}=\eta^{*}$ also
decreases. By these arguments, we have an explanation for the diverging singularity of the
shear viscosity at the phase transition in the hexagonal geometry.

This results show that, at a fluid-solid phase transition the viscosity
coefficients may depend sensitively on the geometry of the lattice of the
solid phase in formation.

\vspace*{0.5cm}

%%%%%%%%%%%%%%%%%%%%%%%%%%%%%%%%%%%%%%%%%%%%%%%%%%%%%%%%%%%%%%%%%%%%%%%%%%%%%%%%

\subsection{Viscosity by the method of Alder {\it et al.}}

We have also verified numerically that our method of calculation of the
viscosity based on the Helfand moment (\ref{helfand-torus}) 
gives the same values as the method of Alder {\it et al.} 
based on the expression (\ref{alder-expression}) \cite{alder}.  
In the two-disk system, this expression reduces to:

\begin{equation}
G_{ij}(t)=\sum_{c}^{} \left[ 2\; \frac{p_{i}
\, p_{j}}{m}\; \Delta t_{c-1,c} +
\Delta p_{i}^{(c)} \: r_{j}(t_{c})
\; \theta (t-t_{c}) \right] .
\end{equation}

\begin{figure}
\centerline{\epsfig{file=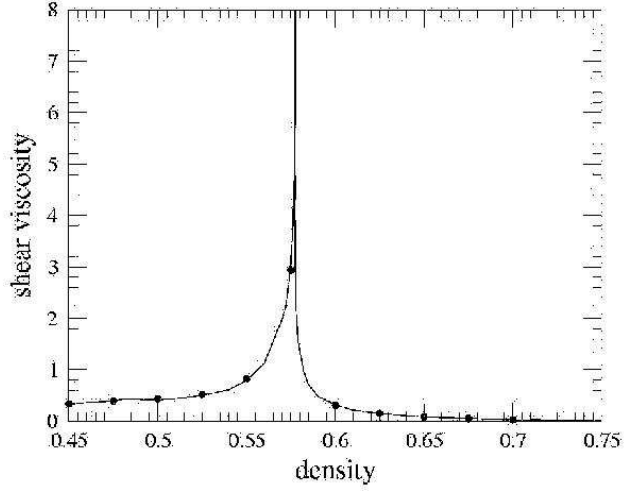,width=9cm}}
\vspace*{0.3cm}
\caption{Shear viscosity $\eta^*$ in the hexagonal geometry calculated by our Helfand
moment (\ref{helfand-torus}) (continuous line) and the one of Alder {\it et al.}
(dots).}
\label{alder-viscosity}
\end{figure}

As shown in Fig. \ref{alder-viscosity} for the shear viscosity in the
hexagonal geometry, there is an
excellent agreement between the values obtained by both methods, which
confirms the exact equivalence
of both methods.

\vspace*{0.5cm}

%%%%%%%%%%%%%%%%%%%%%%%%%%%%%%%%%%%%%%%%%%%%%%%%%%%%%%%%%%%%%%%%%%%%%%%%%%%%%%%%
%%%%%%%%%%%%%%%%%%%%%%%%%%%%%%%%%%%%%%%%%%%%%%%%%%%%%%%%%%%%%%%%%%%%%%%%%%%%%%%%

\section{VISCOSITY IN SYSTEMS OF $N$ HARD DISKS}
\label{sec.Ndisks.visc}

In the present section, we apply our Helfand-moment method to systems of
$N$ hard disks.  Our purpose is to show that the values of the shear
viscosity obtained for the two-hard-disk model are in good agreement
with the values for larger systems, as well as with Enskog's theory.

\begin{figure}
\centerline{\epsfig{file=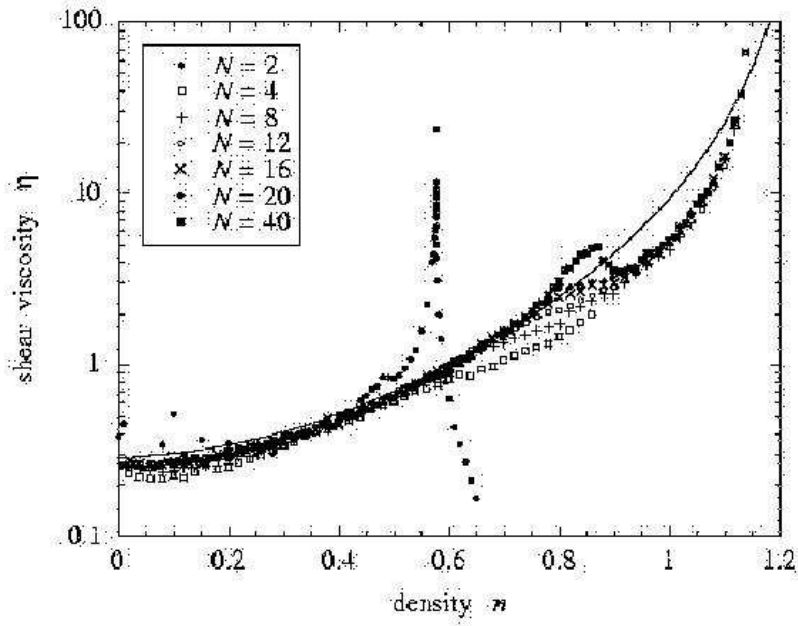,width=12cm}}
\vspace*{0.3cm}
\caption{Shear viscosity $\eta$ versus particle density $n$ in fluids at temperature
$T=1$ with $N=2,4,8,12,16,20,40$ hard disks of unit mass and diameter.  The
solid line is Enskog's value (\ref{Enskog.visc}). For $N=2$, the data are the same as
in Fig. \ref{viscosite-hex} except that we here plot $\eta=2\eta^*$ instead of
$\eta^*$ as in Fig. \ref{viscosite-hex}.}
\label{Ndisk-viscosity}
\end{figure}

Figure \ref{Ndisk-viscosity} depicts the shear viscosity of systems
containing from $N=2$ up to $N=40$ hard disks.  For $N=2$, we consider
here the hexagonal geometry.  For the systems with $N=4$-$40$ disks, the time
evolution is simulated by molecular dynamics with periodic boundary conditions in
the square geometry. The viscosity is calculated by the Helfand-moment method based
on Eq. (\ref{alder-expression}).

We observe that, at low densities, the numerical values are in very good agreement
between themselves. At higher densities, differences appear because the fluid-solid
transition shifts toward higher densities as the number of disks increases.
For $N=2$ disks in the hexagonal geometry, the fluid-solid transition
occurs in the interval $n=0.57$-$0.60$, while it occurs
in the interval $n=0.87$-$0.90$ for $N=40$.  The sharp singularity 
of viscosity for $N=2$ in the hexagonal geometry is specific to the geometrical
constraints of a two-degree-of-freedom system, as explained in the previous section.
Nevertheless, we notice that the decrease of the shear viscosity just above
the fluid-solid transition is also the feature of the large system
with $N=40$ disks.  

Furthermore, the results of our Helfand-moment method are compared with Enskog's
theory.  For a fluid of hard disks of mass $m$ and diameter $\sigma$, Enskog's theory
predicts that the shear viscosity is given by \cite{G71}
\begin{equation}
\eta = \eta_0 \; \left( \frac{1}{Y} + 2\, y + 3.4916 \; Y \; y^2\right) \; ,
\label{Enskog.visc}
\end{equation}
where
\begin{equation}
\eta_0 =  \frac{1.022}{2 \; \sigma} \sqrt{\frac{m\; k_{\rm B}T}{\pi}} \; .
\end{equation}
is the Boltzmann value of the shear viscosity, $Y$ is the Enskog factor entering the
equation of state as follows:
\begin{equation}
P = nk_{\rm B}T (1+2\, y\; Y) \; .
\end{equation}
and $y=\pi\sigma^2 n/4$.  For the hard-disk fluid, a good approximation of the Enskog
factor is given below the fluid-solid transition by \cite{BH76}:
\begin{equation}
Y = \frac{1-\frac{7}{16}\; y}{(1-y)^2} \; .
\end{equation}
It is known that the Enskog approximation is not good around the fluid-solid
transition and at very high densities.

A remark is here in order.  It is known \cite{tail} that the viscosity coefficient
of the infinite hard-disk fluid is diverging because of long-time tails.
However, this divergence is only logarithmic and does not manifest itself
in numerical calculations before extremely long times.  This explains why
the long-time tails do not spoil the agreement between the numerical
values and Enskog's theory.

We see in Fig. \ref{Ndisk-viscosity} the good agreement between Enskog's theory
and the numerical values of our Helfand-moment method at low densities showing
the consistency of our results.

%%%%%%%%%%%%%%%%%%%%%%%%%%%%%%%%%%%%%%%%%%%%%%%%%%%%%%%%%%%%%%%%%%%%%%%%%%%%%%%%
%%%%%%%%%%%%%%%%%%%%%%%%%%%%%%%%%%%%%%%%%%%%%%%%%%%%%%%%%%%%%%%%%%%%%%%%%%%%%%%%

\section{Conclusions}
\label{sec.conclusions}

In this paper, we propose a new expression for the Helfand moment associated with
viscosity in molecular dynamics with periodic boundary conditions.  This new
Helfand moment takes into account the minimum image convention at the basis of
molecular-dynamics simulations with periodic boundary conditions.  In order to
satisfy the minimum image convention, the position coordinates of the particles
undergo jumps.  These jumps modify both the equations of motion and the Helfand
moment which is given by the time integral of the microscopic current entering the
Green-Kubo formula.  As a consequence, the viscosity tensor calculated with our
Helfand moment is equivalent to the one based on the Green-Kubo formula, as proved
in Appendix \ref{appendixB}.  In the case of hard-ball systems, we also prove in
Subsec. \ref{subsec.Helfand-hard-ball} that our method is equivalent to the method by
Alder {\it et al.} \cite{alder}.  Moreover, we show in Appendix \ref{appendixC}
that the hydrostatic pressure can also be calculated thanks to the Helfand moment we
propose for molecular dynamics with periodic boundary conditions.  Our new Helfand
moment and our proofs bring a solution to the ambiguities and problems reported by
Erpenbeck \cite{erpenbeck} about the definition of a Helfand moment in a molecular
dynamics with periodic boundary conditions.  We think that the Helfand-moment method
can be very useful for the numerical calculation of viscosity because this method
has the advantage of being numerically robust.  

We have applied our Helfand-moment method to the numerical calculation of
viscosity in systems of hard disks.

In Sec. \ref{sec.2disks.visc}, viscosity has been studied in detail in a simple
model composed of two hard disks in elastic collision.  This model has already been
investigated in the square geometry by Bunimovich and Spohn \cite{buni-spohn}.  In
the present paper, we generalize this model to the hexagonal geometry.  First, we
show that the fourth-order viscosity tensor is isotropic in the hexagonal geometry
although it is not in the square geometry.  Secondly, we show the viscosity can be
positive and finite in the fluid phase of the hexagonal geometry, although it is
always infinite in the fluid phase of the square geometry.  The reason is
that the horizon of the Sinai billiard driving the dynamics of the two-disk model
is always infinite in the fluid phase of the square geometry although there is a
regime with a finite horizon in the fluid phase of the hexagonal geometry.  In an
infinite-horizon regime, the viscosity becomes infinite so that, from a physical
point of view, the proof of the existence of a positive and finite viscosity
coefficient strictly holds in the hexagonal two-disk model.  In the solid phase, the
transport coefficients acquire a different meaning because the spontaneous breaking
of translational invariance modifies the structure of the hydrodynamic modes and the
viscosity coefficient should be reinterpreted in terms of the damping coefficients
of the transverse sound modes and of the diffusive modes
\cite{martin,fleming,kirkpatrick}.  We hope to report on this question in a future
publication.

The two-disk model presents a phase transition between a fluid and a solid
phase.  This transition is reminiscent of the fluid-solid transition in the system
composed of many disks.  Indeed, the transition manifests itself in the hydrostatic
pressure in a very similar way as in the many-particle system.  The hydrostatic
pressure can be directly related to the mean free path in the two-disk model and we
can thus explain the manifestation of the transition on the pressure in terms of the
behavior of the mean free path near the transition.  In this simple model, the
transition can be understood as a geometrical property of the dynamical system.
Indeed, the trajectories are unbounded in the fluid phase albeit there remain
localized in bounded domains in the solid phase where ergodicity is broken.  The
fluid-solid transition also manifests itself as a diverging singularity in the
viscosity in the two-disk model. We have here shown that this singularity in the
viscosity versus the density may depend sensitively on the geometry of the lattice
of the solid phase in formation.

In Sec. \ref{sec.Ndisks.visc}, we have extended the calculation of shear viscosity
to systems with many disks.  The remarkable result is that the two-disk systems
already gives the shear viscosity in quantitative agreement with its values
in larger systems, as well as with Enskog's theory at moderate densities.

In a companion paper, we report a study of viscosity by the escape-rate method
\cite{viscardy-gasp-2}.  In this other work, we use the Helfand moment which we have
introduced in the present paper.

%%%%%%%%%%%%%%%%%%%%%%%%%%%%%%%%%%%%%%%%%%%%%%%%%%%%%%%%%%%%%%%%%%%%%%%%%%%%%%%%
\vskip 0.5 cm

{\bf Acknowledgments.}
The authors thank Professors J. R. Dorfman and G. Nicolis for support and
encouragement in this research.  We also thank Dr. I. Claus for helpful discussions.
SV thanks the Fund for the Education to the Research in Industry and Agriculture
(FRIA) for financial support. PG thanks the National Fund for Scientific Research
(FNRS Belgium) for financial support.

%%%%%%%%%%%%%%%%%%%%%%%%%%%%%%%%%%%%%%%%%%%%%%%%%%%%%%%%%%%%%%%%%%%%%%%%%%%%%%%%

\begin{appendix}

\section{Microscopic derivation of the viscosity tensor}
\label{appendixA}

In this Appendix, we provide a short microscopic derivation of the
viscosity tensor.

First, we need the balance equation for the local conservation of momentum.
If we define the density of momentum as
\be
g_i({\bf r}) = \sum_{a=1}^N {p}_{ai} \; \delta({\bf r}-{\bf r}_a) \; ,
\ee
the balance equation is
\be
\partial_t \, g_i + \partial_j \, \tau_{ij} = 0 \; ,
\label{mom.bal}
\ee
with $\partial_j=\partial/\partial r_j$. The microscopic momentum current
density is given by
\be
\tau_{ij} = \sum_{a=1}^N \frac{p_{ai}\, p_{aj}}{m} \; \delta({\bf r}-{\bf r}_a)
+ \frac{1}{2} \sum_{a\neq b = 1}^{N} F_i({\bf r}_a-{\bf r}_b) \int_0^1
d\lambda\;
\frac{dr_{abj}}{d\lambda} \; \delta\left[ {\bf r}-{\bf
r}_{ab}(\lambda)\right] \; ,
\ee
where ${\bf r}_{ab}(\lambda)$ is the parametric equation of a curve joining
the particles $a$ and
$b$: ${\bf r}_{ab}(0)={\bf r}_b$ and ${\bf r}_{ab}(1)={\bf r}_a$.

The microscopic current associated with viscosity is defined by integrating
the momentum current
density over the volume $V$:
\be
J_{ij} = \int_V \tau_{ij}({\bf r}) \; d{\bf r} \; ,
\ee
which is given by Eq. (\ref{flux}).
We notice that the hydrostatic pressure is given at equilibrium by
\be
\langle J_{ij} \rangle_{\rm eq} = P \, V \, \delta_{ij} \; ,
\ee
if second-order tensors are isotropic in the system of interest.

We suppose that, at the initial time, the fluid is close to the equilibrium
and described by the following nonequilibrium distribution:
\be
{\mathcal P}(\Gamma) = {\mathcal P}_{\rm eq}(\Gamma)
\left[ 1 + \beta \int {\bf g}({\bf r}) \cdot {\bf v}({\bf r}) \; d{\bf r}
\right]
= {\mathcal P}_{\rm eq}(\Gamma)
\left[ 1 + \beta \sum_{a=1}^N {\bf p}_a \cdot {\bf v}({\bf r}_a)\right] \; ,
\label{proba.0}
\ee
where ${\mathcal P}_{\rm eq}$ is the equilibrium distribution and $\beta$ is a
normalization constant such that
\be
\langle p_{ai}\; p_{bj}\rangle_{\rm eq} = \frac{m}{\beta} \; \delta_{ij} \;
\delta_{ab} \; .
\ee
In the microcanonical state, we have that
\be
\beta=\frac{1}{k_{\rm B}T}\; \frac{N}{N-1} \; .
\ee
The aforementioned distribution describes a fluid with a macroscopic
velocity field ${\bf v}({\bf
r})$ since the nonequilibrium average of the momentum density can easily be
shown to be given by
\be
\langle {\bf g}({\bf r}) \rangle_{\rm noneq} = \rho_{\rm eq} \, {\bf
v}({\bf r}) \; ,
\ee
where
\be
\rho_{\rm eq} = m \, \frac{N}{V} \; ,
\ee
is the mass density at equilibrium.

The time evolution of the probability density (\ref{proba.0}) is ruled by
Liouville's operator
given by the Poisson bracket with the Hamiltonian $\hat L=\{ H,\cdot\}$ or
the pseudo-Liouville
operator in the case of hard-ball dynamics. This operator has the effect of
replacing the
phase-space coordinates $\Gamma$ by $\Gamma(-t)$
\be
{\cal P}_t = {\rm e}^{\hat Lt}{\cal P}_0 = {\mathcal P}_{\rm eq}(\Gamma)
\left[ 1 + \beta \int
{\rm e}^{\hat Lt}{\bf g}({\bf r}) \cdot {\bf v}({\bf r}) \; d{\bf r} \right]
= {\mathcal P}_{\rm eq}(\Gamma) \left[ 1 + \beta
\sum_{a=1}^N {\bf p}_a(-t) \cdot {\bf v}\left[{\bf r}_a(-t)\right]\right] \; .
\label{proba.t}
\ee
Alternatively, we known that the time evolution of the momentum density is
given by
Eq. (\ref{mom.bal}).  In this case, the momentum density should be considered
as an observable so that the solution of Eq. (\ref{mom.bal}) is
\be
{\bf g}({\bf r},t) = {\rm e}^{-\hat Lt}{\bf g}({\bf r},0) \; ,
\ee
so that
\be
{\rm e}^{\hat Lt}{\bf g}({\bf r}) = {\bf g}({\bf r},-t) \; ,
\ee
is solution of the equation
\be
\partial_t \, g_i = \partial_j \, \tau_{ij} \; .
\ee
Integrating both sides over time we get
\be
g_i({\bf r},-t) = g_i({\bf r},0) + \int_0^t dt' \; \partial_j \,
\tau_{ij}(t') \; .
\label{integral}
\ee

Close to equilibrium, we may consider the time evolution of deviations
with respect to the equilibrium. We neglect terms which are quadratic in
the deviations such as the velocity field itself.  The time evolution of these
deviations is obtained by considering the nonequilibrium average of the balance
equation (\ref{mom.bal})
for the deviations:
\be
\partial_t \, \langle\delta g_i\rangle_{\rm noneq} + \partial_j \,
\langle\delta\tau_{ij}\rangle_{\rm noneq} =0 \; ,
\ee
with
\be
\delta \tau_{ij} = \tau_{ij} - \langle\tau_{ij}\rangle_{\rm eq} \; .
\ee
The nonequilibrium average of the deviation of the momentum current density
is given by
\be
\langle \delta\tau_{ij}({\bf r})\rangle_{\rm noneq} = \int
\delta\tau_{ij}({\bf r}) \;
{\cal P}(\Gamma,t)\;  d\Gamma = \beta
\int d{\bf r}'  \langle \delta\tau_{ij}({\bf r}) \; g_k({\bf r}',-t)
\rangle_{\rm eq} \; v_k({\bf
r}') \; .
\label{pre}
\ee
We use Eq. (\ref{integral}) to transform the average as
\be
\langle \delta\tau_{ij}({\bf r}) \; g_k({\bf r}',-t) \rangle_{\rm eq}
=
\langle \delta\tau_{ij}({\bf r}) \; g_k({\bf r}',0) \rangle_{\rm eq}
+ \int_0^t dt' \; \langle \delta\tau_{ij}({\bf r},0)\;
\partial_l' \, \delta \tau_{kl}({\bf r}',t') \rangle_{\rm eq} \; ,
\label{id}
\ee
where we have used the property that $\partial_l' \langle
\tau_{kl}\rangle_{\rm eq}=0$
because the equilibrium state is spatially uniform.  We notice that the first
term in the right-hand side of Eq. (\ref{id}) vanishes because the equilibrium
average of an odd power of particle momenta vanishes.  After an integration
by part
over the velocity field, Eq. (\ref{pre}) becomes
\be
\langle \delta\tau_{ij}({\bf r})\rangle_{\rm noneq} = - \beta \int d{\bf r}'
\int_0^t dt'\; \langle \delta\tau_{ij}({\bf r},0) \;
\delta \tau_{kl}({\bf r}',t') \rangle_{\rm eq} \;
\partial_l'v_k({\bf r}')  = -\eta_{ij,kl}\; \partial_lv_k({\bf r}) \; ,
\ee
where the identification with the viscosity tensor is carried out in the
limit $t\to\infty$
by
\be
\eta_{ij,kl} \; \delta({\bf r}-{\bf r}') =\beta
\int_0^{\infty} dt'\; \langle \delta\tau_{ij}({\bf r},0) \;
\delta \tau_{kl}({\bf r}',t') \rangle_{\rm eq} \; .
\label{eta-delta}
\ee
Taking the double volume integral $\int_V d{\bf r} \int_V d{\bf r}'$ of
both sides of Eq.
(\ref{eta-delta}) and dividing by the volume $V$, we obtain the viscosity
tensor as
\be
\eta_{ij,kl}  = \frac{\beta}{V}
 \int_0^{\infty} dt \; \langle \delta J_{ij}(0) \;
\delta J_{kl}(t) \rangle_{\rm eq} \; ,
\ee
with
\be
\delta J_{ij}(t) = \int_V d{\bf r} \; \delta\tau_{ij}({\bf r},t) =
J_{ij}(t) - \langle
J_{ij}\rangle_{\rm eq} \; ,
\label{dJ}
\ee
Q.E.D.

\section{Proof of the equivalence between Green-Kubo and Einstein-Helfand formulas}
\label{appendixB}

Our aim is here to deduce the Green-Kubo formula (\ref{autocorrel}) from the
Einstein-Helfand formula (\ref{Einstein}), proving the equivalence between both
formulas under the condition that the Helfand moment is defined by Eq.
(\ref{helfand-integral}) as the time integral of the microscopic current (\ref{flux})
and the further condition that the time auto-correlation functions decrease fast
enough.

We start from the Einstein-Helfand formula (\ref{Einstein}) with
\begin{equation}
\delta G_{ij}(t) = \int_0^t \delta J_{ij}(\tau) \; d\tau \; ,
\end{equation}
$\delta J_{ij}$ being defined by Eq. (\ref{dJ}) and supposing for simplicity that
$\delta G_{ij}(0)=0$.  Accordingly, we have successively from Eq. (\ref{Einstein})
that
\begin{eqnarray}
\eta_{ij,kl} &=& \lim_{T\to\infty} \frac{\beta}{2TV} \; \langle \delta G_{ij}(T)
\delta G_{kl}(T) \rangle \nonumber\\
&=& \lim_{T\to\infty} \frac{\beta}{2TV} \; \int_0^T dt_1 \int_0^T dt_2 \;
\langle \delta J_{ij}(t_1) \delta J_{kl}(t_2) \rangle \nonumber\\
&=& \lim_{T\to\infty} \frac{\beta}{2TV} \; \int_{-T}^{+T} dt 
\int_{\vert t\vert/2}^{T-\vert t\vert/2} d\tau
\; \langle \delta J_{ij}(0) \delta J_{kl}(t) \rangle \nonumber\\
&=& \lim_{T\to\infty} \frac{\beta}{2V} \; \int_{-T}^{+T} dt 
\left( 1 - \frac{\vert t\vert}{T}\right) 
\; \langle \delta J_{ij}(0) \delta J_{kl}(t) \rangle \nonumber\\
&=&  \frac{\beta}{2V} \; \int_{-\infty}^{+\infty} dt 
\; \langle \delta J_{ij}(0) \delta J_{kl}(t) \rangle \nonumber\\
&=&  \frac{\beta}{V} \; \int_{0}^{+\infty} dt 
\; \langle \delta J_{ij}(0) \delta J_{kl}(t) \rangle \; ,
\end{eqnarray}
where we have performed the change of integration variables
\begin{equation}
\cases{ t = t_2-t_1 \; , \cr \tau = \frac{t_1+t_2}{2} \; ,\cr}
\end{equation}
and supposed that
\begin{equation}
\lim_{T\to\infty} \frac{1}{T} \; \int_{-T}^{+T} dt \;
\vert t\vert \; \langle \delta J_{ij}(0) \delta J_{kl}(t) \rangle = 0 \; ,
\end{equation}
which requires that the time autocorrelation functions decrease faster than
$\vert t \vert^{-1-\epsilon}$ with $\epsilon>0$. Q.E.D.

\section{Pressure and Helfand moment}
\label{appendixC}

The hydrostatic pressure at equilibrium is given as the mean value of the
momentum current density, i.e., as the mean value of the same microscopic current
entering the Green-Kubo relation:
\be
P_{ij} V = \int_V \langle \tau_{ij}\rangle_{\rm eq} \; d{\bf r} = \langle
J_{ij}\rangle_{\rm eq} \; .
\ee
The average over the thermodynamic equilibrium state can be replaced by a
time average:
\be
P_{ij} V = \langle J_{ij}\rangle_{\rm eq} = \lim_{t\to\infty} \frac{1}{t}
\int_0^t d\tau \; J_{ij} \; .
\ee
We can here introduce the Helfand moment to obtain the hydrostatic pressure
from the Helfand moment as:
\be
P_{ij} V = \lim_{t\to\infty} \frac{1}{t} \left[ G_{ij}(t)-G_{ij}(0)\right] \; .
\ee
In the microcanonical equilibrium state we have that
\be
\langle p_{ai}\; p_{aj}\rangle_{\rm eq} = m \; k_{\rm B}T \;
\frac{N-1}{N}\; \delta_{ij} \; .
\ee
If we assume that the system is isotropic, $P_{ij}=P\; \delta_{ij}$ and we
obtain
\be
PV = (N-1)k_{\rm B}T + R \; ,
\ee
where the rest $R$ provides the corrections to the law of perfect gases in
dense systems. By using
Eqs. (\ref{helfand-torus}) and (\ref{alder-expression}), the virial can be
computed alternatively by
\begin{eqnarray}
R &=& \left\langle \frac{1}{2d} \sum_{a\neq b=1}^{N} {\bf F}({\bf
r}_{ab})\cdot {\bf r}_{ab}
\right\rangle_{\rm eq}\\
&=& \lim_{t\to\infty} \frac{-1}{td} \sum_s\sum_{a=1}^N {\bf
p}_{a}^{(s)}\cdot \Delta{\bf
r}_{a}^{(s)} \; \theta(t-t_s)\\
&=& \lim_{t\to\infty} \frac{1}{td} \sum_c \Delta{\bf p}_{a}^{(c)}\cdot
{\bf r}_{ab}^{(c)} \;
\theta(t-t_c)
\label{rest-coll}
\end{eqnarray}
where $d$ is the dimension, ${\bf r}_{ab}={\bf r}_a-{\bf r}_b$, $t_s$ are
the times of
jumps to satisfy the minimum image convention, while the last expression
only holds for hard-ball
systems, $t_c$ are the collision times, and ${\bf r}_{ab}^{(c)}={\bf
r}_a(t_c)-{\bf r}_b(t_c)$.

\end{appendix}

%%%%%%%%%%%%%%%%%%%%%%%%%%%%%%%%%%%%%%%%%%%%%%%%%%%%%%%%%%%%%%%%%%%%%%%%%%%%%%%%

\end{document}